\documentclass[12pt]{article}
 
\global\arraycolsep=1pt
\usepackage{graphicx}
\usepackage{amssymb}
\usepackage {pstricks}
\newcommand{\beq}{\begin{equation}}
\newcommand{\eeq}{\end{equation}}
\newcommand{\beqa}{\begin{eqnarray}}
\newcommand{\eeqa}{\end{eqnarray}}
\newcommand{\CR}{\nonumber \\}
\newcommand{\m}{\mu}
\newcommand{\n}{\nu}
\renewcommand{\k}{\kappa}

\renewcommand{\theequation}{\thesection.\arabic{equation}}
\renewcommand{\thefootnote}{\fnsymbol{footnote}}

\newcommand{\unitbox}
{\setlength{\unitlength}{0.5pt}
\begin{picture}(10,10)
\put(0,10){\line(1,0){10}}
\put(0,0){\line(1,0){10}}
\put(0,0){\line(0,1){10}}
\put(10,0){\line(0,1){10}}
\end{picture}}

\begin{document}

\begin{titlepage}
\begin{flushright}
{October, 2003}\\
{preprint UT 03-35}\\
{\tt hep-th/0310235} \\
\end{flushright}
\vspace{0.5cm}
\begin{center}
{\Large \bf
Topological Strings and Nekrasov's formulas }
\vskip1.0cm
{Tohru Eguchi}
\vskip 1.0em
{\it Department of Physics, Faculty of Science \\
University of Tokyo, Tokyo, 113-0033, Japan}
\vskip 0.8cm
{ Hiroaki Kanno}
\vskip 1.0em
{\it 
Graduate School of Mathematics \\
Nagoya University, Nagoya, 464-8602, Japan}
\end{center}
\vskip0.5cm

\begin{abstract}
We apply the method of geometric transition and compute all genus
topological closed string amplitudes
compactified on local ${\bf F}_0$
by making use of the Chern-Simons gauge theory. We find an exact agreement of the 
results of our computation with the formula
proposed recently by Nekrasov for ${\cal N}=2$ $SU(2)$ gauge theory with two  
parameters $\beta$ and $\hbar$. $\beta$ is related to the size of the fiber of ${\bf F}_0$
and $\hbar$ corresponds to the string coupling constant.

Thus Nekrasov's formula encodes all the information of topological string 
amplitudes on local ${\bf F}_0$ 
including the number of holomorphic curves at arbitrary genus.
By taking suitable limits $\beta$ and/or $\hbar \rightarrow 0$ one recovers the
four-dimensional Seiberg-Witten theory and
also its coupling to external graviphoton fields.

We also compute topological string amplitude for the local 2nd del Pezzo surface and check
the consistency with Nekrasov's formula of $SU(2)$ gauge theory with a matter field
in the vector representation.

\end{abstract}
\end{titlepage}


\renewcommand{\thefootnote}{\arabic{footnote}}
\setcounter{footnote}{0}

\section{Introduction}
\setcounter{equation}{0}

Recently the method of geometric transition has been applied to various cases of local 
Calabi-Yau manifolds $M$ and the topological amplitudes of closed string theory 
compactified on $M$ have been computed using the Chern-Simons theory 
\cite{AMV,DFG,Iqb,IK-P1,IK-P2,DF}.
In particular the cases of Calabi-Yau manifolds which are 
canonical bundles $K_S$ over rational surfaces $S$ such as ${\bf P}^2$, Hirzebruch 
surfaces ${\bf F}_n$ and also del Pezzo surfaces have been studied in detail.
It has been shown that the structure of Gopakumar and Vafa \cite{GV} for the number of 
higher genus holomorphic  curves emerges automatically from the all genus Chern-Simons calculations
and currently the method of geometric transition is the most efficient way of
computing higher genus amplitudes of topological strings.

On the other hand, using the technique of localization theory Nekrasov \cite{NekSW}
has recently proposed a
formula for the instanton amplitudes of ${\cal N}=2$ SUSY gauge theory at arbitrary instanton 
numbers: his formula contains
two parameters $\beta$ and $\hbar$ and involves the sum over a pair of Young diagrams
in the case of $SU(2)$ gauge group. In the limit $\beta,\hbar\rightarrow 0$
Nekrasov's expression reduces to the well-known results of Seiberg and Witten of four-dimensional
gauge theory \cite{SW}. See also \cite{LMN, NO, NY} for recent studies on Nekrasov's formula.

In \cite{IK-P1} Iqbal and Kashani-Poor used the method of geometric transition 
for the local ${\bf F}_0$ and
reproduced Nekrasov's expression 
by postulating some identities involving Hopf-link invariants in Chern-Simons 
theory. They were, however, mainly concerned  with the case of 
field theory limit $\beta,\hbar\rightarrow 0$.

In this paper we prove these identities and compute rigorously the all genus
topological string amplitudes on local ${\bf F}_0$.
We find an exact agreement of our results with the formula of Nekrasov with 
parameters $\beta,\hbar$ kept being non-zero: these parameters match with the 
corresponding ones of Chern-Simons theory.
We note that when the parameter $\beta$ is identified as the 
radius $R$ of the fifth-dimensional circle, Nekrasov's expression exactly reproduces the instanton 
amplitudes of five-dimensional gauge theory on ${\bf R}^4\times S^1$ at $\hbar=0$. 
We recall that instanton 
amplitudes in five-dimensional gauge theories count the number of genus zero holomorphic curves
in Calabi-Yau manifolds \cite{Nek5D,LN,KO,EK}.
 
Agreement with the all genus amplitudes obtained by the Chern-Simons calculation
implies that the higher order terms in $\hbar$ of Nekrasov's formula
count precisely the number of higher genus holomorphic curves. Therefore
Nekrasov's expression encodes all the information 
of closed topological string amplitudes compactified 
on the local Calabi-Yau manifold $K_{{\bf F}_0}$.

If one takes the four-dimensional limit 
$\beta\rightarrow 0$ with $\hbar$ kept finite, one obtains the amplitudes of
the four-dimensional 
${\cal N}=2$ gauge theory in the graviphoton background as suggested by \cite{NekSW}.

In the following we will provide proofs for some of the identities involving the Hopf-link 
invariants 
postulated in previous papers.   
We also study the local del Pezzo surfaces which are obtained by blowing up local Hirzebruch surfaces and describe gauge theories with additional matter hypermultiplets. 
We compute the topological string amplitude for the 2nd del Pezzo
surface and recover Nekrasov's expression of $SU(2)$ gauge theory in the presence of
a matter in the vector representation.   

\bigskip


\section{Toric diagram and topological string amplitudes}
\setcounter{equation}{0}

In the following we will 
consider a class of toric rational surfaces $S$ that includes 
${\bf P}^2$, ${\bf F}_0 = {\bf P}^1 \times {\bf P}^1$, 
${\bf F}_2$ and their blow ups. The canonical line bundle $K_S$ over $S$ is
a non-compact Calabi-Yau manifold. 
We will compute topological closed string amplitudes compactified on 
the manifolds $K_S$.
In general toric diagrams describe patterns of degeneration of the torus action.
For a toric rational surface $S$ with ${\bf T}^2$ action, 
the diagram is made of a polygon with $N$ vertices and an external line 
attached to each trivalent vertex. Along the edges of the polygon
an $S^1$ action specified by the direction of each edge degenerates 
and at the vertices both of the ${\bf T}^2$ actions become degenerate.
The diagram looks like a one loop Feynman graph with $N$ external lines,
where $N=3$ for ${\bf P}^2$, $N=4$ for ${\bf F}_m$ and 
blowing up at a point increases $N$ by one.
We take the clockwise direction 
in the loop and assign to the $i$-th internal edge 
a representation $R_i~(i=1, \cdots, N)$ and the (renormalized) 
K\"ahler parameter $t_i$. Note that by the way of degeneration of toric
action explained above, we have a rational curve for each internal edge.
All external lines have the trivial representation that will be 
denoted by $\bullet$. Throughout the paper 
we will identify representation $R$ with its Young diagram $\m^R$, 
specified by the number of boxes $\m_j$ in the $j$-th row. 
The $\m_j$'s are non-increasing; $\m_1 \geq \m_2
\geq \cdots \geq \m_d > \m_{d+1}=0$, where $d:= d(\m^R)$ is 
the \lq\lq depth\rq\rq\, i.e. the number of rows of the diagram. 
We denote the total number of boxes by $\ell_R$;
\beq
\ell_R := \sum_{j=1}^d \m_j~.
\eeq
The Young diagram represents a partition of $\ell_R$ objects.
The integer $\k_R$ defined by
\beq
\k_R := 2 \sum_{j=1}^d \sum_{k=1}^{\m_j} (k-j) 
= \ell_R + \sum_{j=1}^d \m_j(\m_j -2j)
\eeq
plays also an important role in the following. It is related to the quadratic Casimir 
of the representation $R$ of the group $U(N)$ by
\beq
C_R = \k_R + N \ell_R~.
\eeq


\begin{center}

\begin{pspicture}(-7,-2)(5,5) 

\psline[arrowsize=5pt]{->}(-6.5,0.5)(-6.5,1.5)
\psline(-6.5,1.4)(-6.5,2.5)
\psline[arrowsize=5pt]{->}(-6.5,2.5)(-5.3,1.3)
\psline(-5.4,1.4)(-4.5,0.5)
\psline[arrowsize=5pt]{->}(-4.5,0.5)(-5.6,0.5)
\psline(-5.5,0.5)(-6.5,0.5)
\psline(-6.5,0.5)(-7,0)
\psline(-6.5,2.5)(-7,3.5)
\psline(-4.5,0.5)(-3.5,0)
\rput(-6.2,2.6){$P$}
\rput(-6.8,1.4){$H$}
\rput(-5.2,1.7){$H$}
\rput(-5.5,0.2){$H$}
\rput(-5.5,-1){$H\cdot H = +1$}

\psline[arrowsize=5pt]{->}(1,0.5)(1,1.6)
\psline(1,1.5)(1,2.5)
\psline[arrowsize=5pt]{->}(1,2.5)(1.7,2.5)
\psline(1.6,2.5)(2,2.5)
\psline[arrowsize=5pt]{->}(2,2.5)(3,1.5)
\psline(2.9,1.6)(4,0.5)
\psline[arrowsize=5pt]{->}(4,0.5)(2.2,0.5)
\psline(2.3,0.5)(1,0.5)
\psline(1,0.5)(0,-0.5)
\psline(1,2.5)(0,3.5)
\psline(2,2.5)(2,3.5)
\psline(4,0.5)(5,0)
\rput(0.6,1.5){$F$}
\rput(3.3,1.7){$F$}
\rput(1.6,2.9){$B$}
\rput(2.5,0.2){$H=B+F$}
\rput(2.5,-1){$B\cdot B = -1~,~F\cdot F =0$}

\end{pspicture}

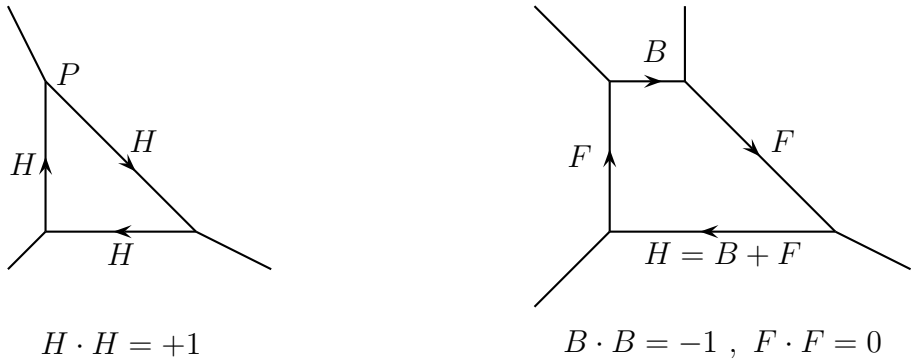
Figure 1 :  Toric diagram of ${\bf P}^2$ (left) and its blow up ${\bf F}_1$ at P (right)

\end{center}


Based on the rules proposed by Iqbal \cite{Iqb} and 
the idea of topological vertex \cite{AKMV}, 
we find the following universal formula for
topological closed string amplitude compactified on local Calabi-Yau 
manifolds $K_S$;
\beqa
Z_{top~str}^{(S)} &=& \sum_{R_1 \cdots R_N} W_{R_N R_1} W_{R_1 R_2} 
\cdots W_{R_{N-1} R_N} \cdot e^{- \sum_{i=1}^N t_i \cdot \ell_{R_i}} \CR
& &~~~\times (-1)^{\sum_{i=1}^N \gamma_i \cdot \ell_{R_i}}
q^{\frac{1}{2}\sum_{i=1}^N \gamma_i \cdot \k_{R_i}}~,
\eeqa
where $q=\exp(2\pi i/(N+k))$ and $W_{R_i R_j}$ is 
the Chern-Simons invariant of the Hopf link (with the standard framing) 
in $S^3$ carrying representations $R_i$ and $R_j$ (see Appendix A). 
The \lq\lq propagator\rq\rq\  for the $i$-th internal line 
is given by $e^{- t_i \cdot \ell_{R_i}}$. The last two factors come from
the choice of the framing. It turns out that we can fix the framing
by the self-intersection number $\gamma_i$
of the rational curve associated with the $i$-th edge. In deriving the above
formula we have used the following relation to the topological vertex 
$C_{R_1, R_2, R_3}$ introduced in \cite{AKMV};
\beq
C_{\bullet, R_2, R_3^t} = W_{R_2 R_3} q^{- \frac{1}{2} \k_{R_3}}~,
\eeq
where $R^t$ denotes the conjugate representation obtained by 
exchanging row and columns of the Young diagram. According to
\cite{AKMV} the topological vertex has the cyclic symmetry;
\beq
C_{R_1, R_2, R_3} = C_{R_2, R_3, R_1} = C_{R_3, R_1, R_2}~,
\eeq
and the conjugation property;
\beq
C_{R_1, R_2, R_3} = q^{\frac{1}{2} \sum_i \k_{R_i}} C_{R_1^t, R_3^t, R_2^t}~.
\eeq
In particular the last property implies;
\beq
W_{R_1 R_2} = q^{\frac{1}{2} \k_{R_2}} C_{\bullet, R_1, R_2^t} 
= q^{\frac{1}{2} \k_{R_1}} C_{\bullet, R_2, R_1^t}  = W_{R_2 R_1}~,
\eeq
where we have used $\k_{R^t} = - \k_R$. 
Let us consider ${\bf P}^2$ as an example. The toric diagram is 
a triangle ($N=3$) and all the edges represent the hyperplane 
class $H$ of ${\bf P}^2$ with the self-intersection $H \cdot H =1$.
Thus our universal formula implies
\beqa
Z_{top~str}^{({\bf P}^2)} &=& \sum_{R_1, R_2, R_3} W_{R_1 R_2} W_{R_2 R_3} W_{R_3 R_1} 
\cdot e^{- t_H \cdot (\ell_{R_1} + \ell_{R_2} +\ell_{R_3})} \CR
& &~~~\times (-1)^{\ell_{R_1} + \ell_{R_2} +\ell_{R_3}}
q^{\frac{1}{2}(\k_{R_1} + \k_{R_2} + \k_{R_3})}~,
\eeqa
where $t_H$ is the K\"ahler moduli for $H$.
This topological string amplitude has been extensively 
investigated in \cite{AMV}.

\bigskip

\section{Local Hirzebruch surface and five dimensional gauge theory}
\setcounter{equation}{0}

The Hirzebruch surface ${\bf F}_m$ is a ${\bf P}^1$ bundle over ${\bf P}^1$.
The second homology class $H_2({\bf F}_m, {\bf Z})$ is spanned by
the two cycles $B$ and $F$, where their representatives are
the base ${\bf P}^1$ and the ${\bf P}^1$ fiber, respectively.
The intersection numbers of these cycles are
\beq
B \cdot B = -m~, \quad F \cdot F = 0~, \quad B \cdot F = +1~.
\eeq
The Hirzebruch surfaces ${\bf F}_m~(m=0,1,2)$ have been used in
geometric engineering of ${\cal N}=2$ pure Yang-Mills theory
with $SU(2)$ gauge symmetry \cite{KKV,KMV,KV}.


\begin{center}

\begin{pspicture}(-7,-2)(5,5) 

\psline[arrowsize=5pt]{->}(-6,0.5)(-6,1.6)
\psline(-6,1.5)(-6,2.5)
\psline[arrowsize=5pt]{->}(-6,2.5)(-4.9,2.5)
\psline(-5,2.5)(-4,2.5)
\psline[arrowsize=5pt]{->}(-4,2.5)(-4,1.4)
\psline(-4,1.5)(-4,0.5)
\psline[arrowsize=5pt]{->}(-4,0.5)(-5.1,0.5)
\psline(-5,0.5)(-6,0.5)
\psline(-6,0.5)(-7,-0.5)
\psline(-6,2.5)(-7,3.5)
\psline(-4,2.5)(-3,3.5)
\psline(-4,0.5)(-3,-0.5)
\rput(-3.7,1.5){$F$}
\rput(-6.3,1.5){$F$}
\rput(-5,0.2){$B$}
\rput(-5,2.8){$B$}
\rput(-5,-1.5){$B\cdot B = 0$~,~$F\cdot F = 0$}

\psline[arrowsize=5pt]{->}(0.5,0.5)(1.5,1.5)
\psline(1.4,1.4)(2,2)
\psline[arrowsize=5pt]{->}(2,2)(2.7,2)
\psline(2.6,2)(3,2)
\psline[arrowsize=5pt]{->}(3,2)(3.8,1.2)
\psline(3.7,1.3)(4.5,0.5)
\psline[arrowsize=5pt]{->}(4.5,0.5)(2.4,0.5)
\psline(2.5,0.5)(0.5,0.5)
\psline(0.5,0.5)(-0.5,0)
\psline(2,2)(2,3)
\psline(3,2)(3,3)
\psline(4.5,0.5)(5.5,0)
\rput(1,1.5){$F$}
\rput(4,1.5){$F$}
\rput(2.5,2.4){$B$}
\rput(2.5,0.2){$B+2F$}
\rput(2.5,-1.5){$B\cdot B = -2~,~F\cdot F =0$}

\end{pspicture}

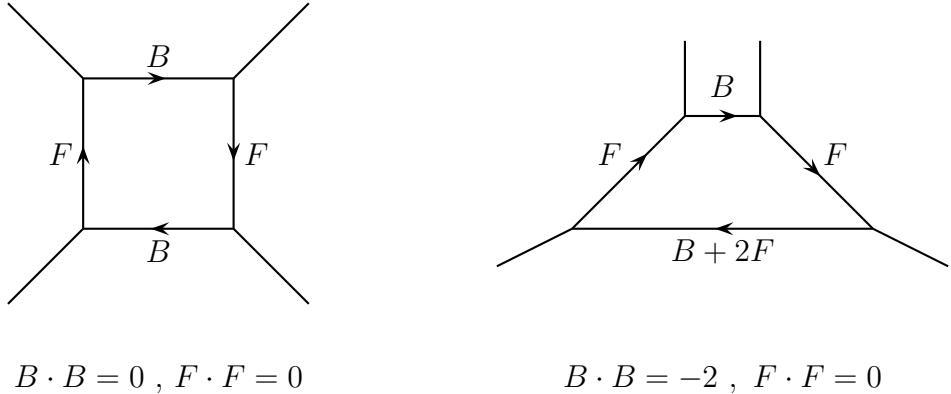
Figure 2 : Toric diagram of the Hirzebruch surfaces ${\bf F}_0$ (left) and ${\bf F}_2$ (right)

\end{center}


From the toric diagram of ${\bf F}_m~(m=0,1,2)$ and the data of self-intersection numbers
we can write down all genus topological string amplitude;
\beqa
Z_{top~str}^{({\bf F}_m)}&=& \sum_{R_1 \cdots R_4} W_{R_4 R_1} W_{R_1 R_2} 
W_{R_2 R_3}  W_{R_3 R_4} 
\cdot e^{- t_F \cdot (\ell_{R_1}+\ell_{R_3} + m \ell_{R_4})
- t_B \cdot (\ell_{R_2}+\ell_{R_4}) } \CR
& &~~~\times (-1)^{m (\ell_{R_4}-\ell_{R_2}) }
q^{\frac{m}{2}(\k_{R_4}-\k_{R_2})}~,
\eeqa
where $t_B$ and $t_F$ are the (renormalized) K\"ahler parameters of $B$ and $F$.
Following \cite{IK-P1}, we introduce the function
\beq
K_{R_1 R_2}(Q) := \sum_S Q^{\ell_S} W_{R_1S}(q) W_{SR_2}(q)~.
\eeq
The amplitude can be written as
\beq
Z_{top~str}^{({\bf F}_m)} = \sum_{R_1 R_2} (K_{R_1 R_2}(Q_F))^2 \cdot 
Q_B^{\ell_{R_1}+\ell_{R_2}} Q_F^{m \ell_{R_2}}
\cdot (-1)^{m (\ell_{R_2}-\ell_{R_1}) }
q^{\frac{m}{2}(\k_{R_2}-\k_{R_1})}~,
\eeq
where $Q_B := e^{-t_B}$ and $Q_F := e^{-t_F}$. Assuming that the contributions
are exhausted by multiple-covers of isolated rational curves, 
Iqbal and Kashani-Poor have proposed the following proposition;\\
{\bf Proposition 1}
\beq
K_{R_1 R_2}(Q) =W_{R_1}(q)W_{R_2}(q)
\exp \left( \sum_{n=1}^\infty \frac{\widetilde{f}_{R_1R_2}(q^n)}{n}Q^n \right)~. 
\label{K-identitya}
\eeq
Here the function $\widetilde{f}_{R_1R_2}$ is given by
\beqa
&&\widetilde{f}_{R_1 R_2}(q) := W_{\unitbox}^2 (q) 
+ f_{R_1}(q) + f_{R_2}(q) + f_{R_1}(q) f_{R_2}(q) W_{\unitbox}^{-2} (q)~,\\
&&f_R (q):= \sum_{i=1}^d \sum_{k=1}^{\m_i} q^{k-i}={q\over (q-1)}\sum_{i=1}^d
\big(q^{\mu_i-i}-q^{-i}\big),\\
&&W_{\unitbox}(q) := \frac{1}{[1]}= \frac{q^{\frac{1}{2}}}{q-1}~.~
\eeqa

We now present a (somewhat lengthy) proof of the important identity (\ref{K-identitya}): 
we start from the formula of
the Hopf link invariants (see Appendix A);
\beq
W_{R_1 R_2} (q) = W_{R_1}(q) q^{\frac{\ell_{R_2}}{2}} s_{\m^{R_2}} ( E_{\m^{R_1}} )~.
\label{Hopfa}\eeq
Here $s_{\m^{R_2}}$ denotes the Schur function for the representation $R_2$ defined by the 
Jacobi-Trudy formula
\beq
s_{\m^{R_2}}=\det(e_i^{\mu_i'-i+j})~,
\eeq
where $\mu^{\prime}= \m^{{R_2}^t}$ denotes the $\mu$-parameters of the dual Young diagram. 
$\{e_i\}$ are elementary symmetric polynomials of the basic variables $x_i$  
\beq
e_i=\sum_{j_1<j_2<\cdots<j_i}x_{j_1}x_{j_2}\cdots x_{j_i}~.
\label{elementary}\eeq
In terms of $\{x_i\}$ the Schur function is given by
\beq
s_{\m^R}(x_i)={\det \left( x_i^{\mu_j+n-j} \right) \over \det \left(x_i^{n-j} \right)}~.
\eeq 
In evaluating
(\ref{Hopfa}) we substitute the values ${e_i^{\,R_1}}$
defined by the expansion 
\beq
E_{\m^{R_1}}(t) = \sum_{i=0}^\infty e_i^{\,R_1} (q) t^i~,
\eeq
into $e_i$.
We will see that ${e_i}$ defined by this expansion in fact
agrees with the expression (\ref{elementary}) when we suitably specialize the
values of $x_i$.

From the definition (\ref{Emu}) of $E_{\m^{R}}(t)$, one easily finds  
\beq
e_1^R (q)={(q-1)\over q}\widetilde{f}_R(q)~,
\label{E1}\eeq
where
\beq
\widetilde{f}_R(q) := f_R(q) + W_{\unitbox}^2(q)~.
\eeq
We note the factorization property of $\widetilde{f}_{R_1R_2}(q)$
\beq
\widetilde{f}_{R_1R_2}(q)
= W_{\unitbox}^{-2}(q)\widetilde{f}_{R_1}(q) \widetilde{f}_{R_2}(q)
=q\,e_1^{R_1}(q)e_1^{R_2}(q)~.
\eeq

Next let us introduce the character $\chi^{(\mu)}$ of a representation $\mu$ 
of the symmetric group and let $\chi^{(\mu)}_{(\vec{j})}$ be its value on the conjugacy 
class $(\vec{j})=(1^{n_1},2^{n_2},\cdot \cdot,j^{n_j},\cdots)$.
The Frobenius formula for the character is given by
\beq
\prod_j( p_j (x_i))^{n_j}\det(x_i^{n-k})=\sum_{\mu}\chi^{(\mu)}_{(\vec{j})}\det(x_i^{\mu_k+n-k})~,
\label{Frobenius}\eeq
where 
\beq
p_j(x_i) := \sum_i {x_i}^j~,
\eeq
is the power sum. (\ref{Frobenius}) may be rewritten as
\beq
\sum_{\mu}\chi^{(\mu)}_{(\vec{j})}s_{\mu}(x_i)=\prod_j( p_j (x_i))^{n_j}~.
\label{Schur-winding}\eeq
We may identify the right-hand-side of (\ref{Schur-winding}) as
Schur functions in the winding basis
\beq
s_{\vec{j}}(x_i)=\prod_j( p_j (x_i))^{n_j}~.
\eeq
Thus in the winding basis Schur functions become monomials in the power sum. 
Now from (\ref{E1}) we find the following relation
\beqa
&&e_1^R(q)=\sum_{i=1}^d \Big(q^{\mu_i-i}-q^{-i}\Big)+{1\over (q-1)}\nonumber \\
&&=\sum_{i=1}^{\infty}
\Big(q^{\mu_i-i}-q^{-i}\Big)+{1\over (q-1)}=\sum_{i=1}^{\infty}q^{\mu_i-i}~.
\eeqa
Thus the function $e_1^R(q)$ is given by the sum of variables $x_i$ specialized at
\beq
x_i = q^{\m_i - i}~, \hskip4mm i=1,2,3,\cdots~.
\label{specialize}\eeq 
General power sum is similarly given by
\beq
p_j( x_i = q^{\m_i - i}) = e_1^R (q^j)~. \label{specify}
\label{ep-rel}\eeq
Note that if we use the standard relation between the generating functions of 
$e_j(x_i)$ and $p_j(x_i)$, we find
\beq
\sum_{j=0}^\infty e_j(x_i) t^j = \exp \left( \sum_{j=1}^\infty \frac {(-1)^{j+1}}{j} p_j(x_i) 
t^j \right)=\exp \left( \sum_{j=1}^\infty \frac {(-1)^{j+1}}{j} e_1(q^j) 
t^j \right)~.
\eeq
Thus $e_j(q)$ for any $j$ can be expressed in terms of $e_1(q^i),\,(i=1,\cdots,j)$.

Now by using (\ref{ep-rel}) we find
\beq
s_{\vec{j}}(x_i)=\prod_j (e_1(q^j))^{n_j}~.
\label{Schur-e1}
\eeq
By evaluating $K_{R_1R_2}(q)$ in the winding-basis and using (\ref{Schur-e1}) we obtain
\beqa
K_{R_1R_2}(q)&=&  W_{R_1}(q)W_{R_2}(q)\sum_{n_j}\prod_j{1\over j^{n_j}n_j!}
\prod_j (e_1^{R_1}(q^j))^{n_j}\prod_j (e_1^{R_2}(q^j))^{n_j}(qQ)^{\sum jn_j}\nonumber \\
&=& W_{R_1}(q)W_{R_2}(q)\exp\Big(\sum_n {q^n\over n}e_1^{R_1}(q^n)e_1^{R_2}(q^n)Q^n\Big)~.
\eeqa
The normalization factor $\prod {1/j^{n_j}n_j!}$ comes from
the orthogonality relation of the characters
\beq
\sum_{\vec{j}} \frac{1}{\prod_j n_j ! j^{n_j}}
\chi^{(\mu^R)}_{(\vec{j})} \cdot \chi^{(\mu^S)}_{(\vec{j})} = \delta_{R S}~.
\eeq
This is the formula we wanted to prove.

If we set $R_1=R_2=\bullet$ in the above formula, we recover the postulated identity for 
evaluating the Chern-Simons amplitude in the deformed conifold \cite{Iqb};
\beq
\sum_R \big(\dim_q R\big)^2 Q^{\ell_R}=\exp\Big(\sum_j{1\over j}{q^j\over 
(q^j-1)^2}Q^j\Big)~.
\label{conifold}
\eeq
In the following discussions it is convenient to introduce the expansion
coefficients $C_k(R_1, R_2)$ by
\beq
f_{R_1 R_2} (q) =  f_{R_1}(q) + f_{R_2}(q) + (q + q^{-1} - 2) f_{R_1}(q) f_{R_2}(q) 
=\sum_k C_k(R_1, R_2) q^k~. \label{Ck}
\eeq
Then we have
\beq
K_{R_1 R_2}(Q) = W_{R_1}(q) W_{R_2}(q) 
\exp \left( \sum_{n=1}^\infty \frac{W_{\unitbox}^2(q^n)}{n} Q^n\right)
\prod_{k} \left( 1 - q^k Q_F \right)^{-C_k(R_1, R_2)}~.
\eeq

\bigskip


\subsection{Nekrasov's conjecture for five dimensional gauge theory}

We now would like to show the equality of Nekrasov's formula for five dimensional 
gauge theory on ${\bf R}^4 \times S^1$
and all genus topological string amplitude for local toric Calabi-Yau manifold. More precisely, 
we prove that the instanton expansion of the partition function of the pure $SU(2)$ Yang-Mills
theory on ${\bf R}^4 \times S^1$ is {\it exactly} the same as the expansion of all genus topological
string amplitude for local Hirzebruch surface ${\bf F}_0 = {\bf P}^1 \times {\bf P}^1$, where the expansion parameter of topological string is  
an appropriate combination of the K\"ahler parameters of ${\bf F}_0$.
Due to the presence of the framing factor coming from non-trivial self-intersection of the base $B$ and
the fiber $F$, the surfaces ${\bf F}_1$ and ${\bf F}_2$ give different amplitudes
as five dimensional gauge theory. But they have the same four dimensional limit as ${\bf F}_0$.

According to Nekrasov we introduce a pair of representations 
$R_1, R_2$ (of $U(N)$) in the case of $SU(2)$ gauge theory.
The Young tableaux $\m^{R_\ell}$ of the representation $R_\ell$ gives a sequence of 
non-increasing integers;
$\m_{\ell,1} \geq \m_{\ell,2} \geq \cdots \geq \m_{\ell, d(\m^{R_\ell})} > \m_{\ell, d(\m^{R_\ell})+1} =0$,
where $\m_{\ell, i}$ is the number of boxes in the $i$-th row of the Young tableaux $\m^{R_\ell}$ and
$d(\m^{R_\ell})$ is the depth of $\m^{R_\ell}$. Then Nekrasov's conjecture for the $k$-instanton contribution
to the partition function is given by
\beq
Z_k^{(5D)} = \sum_{\ell_{R_1} + \ell_{R_2} = k} \prod_{\ell, m \in \{1,2\}} \prod_{i,j = 1}^\infty
\frac{\sinh R \left( a_{\ell m} + \hbar (\m_{\ell,i} - \m_{m,j} + j -i) \right)}
{\sinh R \left( a_{\ell m} + \hbar ( j -i) \right)}~,
\eeq
where $a_{\ell, m} = a_\ell - a_m$ and we have identified the parameter $\beta$ with the radius 
$R$ of the circle.
For $SU(2)$ gauge theory we have $a_{12} = - a_{21} = 2a$. The partition function is decomposed 
into
two factors; the factor $Z_k^{(5D, 1)}$ with $\ell = m$ and  $Z_k^{(5D, 2)}$ with $\ell \neq m$.
As topological string amplitude they have different origins as we shall see.
Each factor can be simplified as follows (we suppress the restriction $\ell_{R_1} + \ell_{R_2} = k$);
\beqa
Z_k^{(5D, 1)} &=&  \prod_{\ell = 1,2} \prod_{i \neq j}^\infty 
\frac{\sinh R \hbar (\m_{\ell,i} - \m_{\ell,j} + j -i)}
{\sinh R \hbar ( j -i) }~, \CR
&=&  \prod_{\ell = 1,2} \prod_{1 \leq i < j < \infty}
\frac{\sinh^2 R \hbar (\m_{\ell,i} - \m_{\ell,j} + j -i)}
{\sinh^2 R \hbar ( j -i) }~.
\eeqa
\beqa
Z_k^{(5D, 2)} &=&  \prod_{i,j = 1}^\infty
\frac{\sinh R \left( 2a + \hbar (\m_{1,i} - \m_{2,j} + j -i) \right)}
{\sinh R \left( 2a+ \hbar ( j -i) \right)}
\frac{\sinh R \left( -2a+ \hbar (\m_{2,i} - \m_{1,j} + j -i) \right)}
{\sinh R \left( -2a+ \hbar ( j - i) \right)}~, \CR
&=& \prod_{i,j = 1}^\infty
\frac{\sinh^2 R \left( 2a + \hbar (\m_{1,i} - \m_{2,j} + j -i) \right)}
{\sinh^2 R \left( 2a+ \hbar ( j -i) \right)}~.
\eeqa

As we have shown before, all genus topological string amplitude 
for the local Hirzebruch surface ${\bf F}_m$ $(m=0,1,2)$ is given by
\beqa
Z_{top~str}^{({\bf F}_m)} &=& \exp \left( \sum_{n=1}^\infty \frac{2}{n} W_{\unitbox}^2 (q^n) Q_F^n \right)
\sum_{R_1, R_2} Q_B^{\ell_{R_1} + \ell_{R_2}} Q_F^{m \ell_{R_2}} 
 (-1)^{m({\ell_{R_2} - \ell_{R_1})}} q^{-\frac{m}{2}(\k_{R_1} +\k_{R_2})}  \CR
& &~~~\times W_{R_1}^2(q) W_{R_2^t}^2 (q) \prod_{k} \left( 1 - q^k Q_F \right)^{-2 C_k(R_1, R_2^t)}~, \label{Fm}
\eeqa
where we have made a replacement $R_2 \to R_2^t$, which is immaterial since we sum over the 
representations $R_2$. 
The factor $W_R(q)$ is the quantum dimension of the representation $R$ or the knot invariant
for an unknot in $S^3$ with the representation $R$;
\beqa
W_R(q)=\dim_q R &=& q^{{\kappa_R\over 4}}\prod_{1 \leq i < j \leq d} \frac{[\mu_i -\mu_j + j -i ]}{[j-i]}
\prod_{i=1}^{d}{\prod_{k=1}^{\mu_i}{1\over  [k-i + d]}}~,\nonumber \\
&=&q^{{\kappa_R\over 4}}\prod_{1 \leq i < j <\infty} \frac{[\mu_i -\mu_j + j -i ]}{[j-i]}~.
\eeqa
We note the relation
\beq
W_{R^t} (q) = q^{-\frac{\kappa_R}{2}} W_R(q)~.
\eeq
The first factor in (\ref{Fm}) gives the perturbative one-loop contribution to the prepotential.
The $k$-instanton part is identified as the sum of terms obeying the condition 
$\ell_{R_1} + \ell_{R_2} =k$
as in the Nekrasov's formula. The framing factor depends on the 
self-intersection
numbers of the divisors in ${\bf F}_m$ and it becomes trivial for ${\bf F}_0$. 
We identify the parameters of the gauge theory side as
\beq
q= e^{-2R\hbar}~, \quad Q_F = e^{-4Ra}~.
\eeq
Then using the identities
\beq
\sum_k C_k (R_1, R_2) = \ell_{R_1}  + \ell_{R_2}~, \qquad
\sum_k k C_k (R_1, R_2) = \frac{1}{2} ( \kappa_{R_1} + \kappa_{R_2} )~,
\eeq
we obtain
\beq
\prod_k \left( 1 - q^k Q_F \right)^{-2 C_k(R_1, R_2^t)} = (4Q_F)^{-\ell_{R_1} - \ell_{R_2}} 
q^{-\frac{1}{2}(\kappa_{R_1} - \kappa_{R_2})} \prod_k 
\frac{1}{\left[ \sinh R (2a + \hbar k) \right]^{2C_k(R_1, R_2^t)}}~,
\eeq
where $\kappa_{R} = - \kappa_{R^t}$.

Establishing the equality $Z_k^{(5D)}$ and $Z_{top~str,k}^{({\bf F}_m)}$
then boils down to the proof of the following proposition;

{\bf Proposition 2}
\beq
\prod_k \frac{1}{ \sinh R (2a + \hbar k)^{C_k(R_1, R_2^t)}} 
= \prod_{i,j=1}^\infty \frac{ \sinh R \left( 2a + \hbar(\m_{1,i} - \m_{2,j} + j - i) \right)}
{\sinh R \left( 2a + \hbar ( j- i) \right) }
\eeq
This identity has been postulated by Iqbal and Kashani-Poor. We present a proof in the
Appendix B.

The $\kappa$ factors arising in the calculation cancel completely in the case of
${\bf F}_0$ we find a complete agreement between Nekrasov's expression and all genus
topological string amplitude on ${\bf F}_0$.
The $\Lambda$ parameter of gauge theory is identified as
\beq
(R\Lambda)^4 = \frac{Q_B}{16 Q_F}~.
\eeq

We can write down the analogues of Nekrasov's formula corresponding to string theory
compactified on ${\bf F}_1$ and ${\bf F}_2$. These formulas contain some extra framing factors 
which become trivial in the four-dimensional limit $q\rightarrow  1$.

In all known cases of local Calabi-Yau manifolds, such as local ${\bf P}^2$ and ${\bf F}_m$,
Chern-Simons calculations automatically
lead to the form of the Gopakumar-Vafa invariants for the topological string amplitude.
In the case of the canonical bundle over Hirzebruch surfaces ${\bf F}_m$, for instance, its 
free-energy has an expansion
\beq
{\cal F}_{top\hskip1mm str}
=\sum_{n,m=0}^{\infty}\sum_{g=0}^{\infty}\sum_{k=1}^{\infty}{N_{n,m}^g\over k(2\sin k{g_s
\over 2})^{2-2g}}
e^{-k(nt_B+mt_F)}~,
\label{GV-inv}\eeq
where $g_s$ denotes the string coupling constant (related to $q$ as $q=e^{ig_s}$)
and $N_{n,m}^g$ are the Gopakumar-Vafa invariants 
computing 
the number of BPS states obtained by M2-branes wrapped around the two cycles $nB+mF$.
Gopakumar-Vafa(GV) invariants have a simple relation to the standard Gromov-Witten(GW) invariants.
In examples GV invariants calculated from Chern-Simons theory are in precise agreement 
with the known GW invariants calculated by using mirror map and holomorphic anomaly, see for instance \cite{Hos,JN}. 

What is striking in the method of geometric transition and Nekrasov's formula
is that they manage to evaluate the sum over $m$ (winding number around the fiber $F$) in 
(\ref{GV-inv}) and the multiple-cover
factor $k$ exactly in a very efficient manner. Only the sum over $n$ (winding around the base
$B$) is left as the sum over the space-time instanton numbers.

We note that the agreement of Nekrasov's formula with the Chern-Simons computation 
implies that Nekrasov's formula encodes the entire information of topological string
amplitudes compactified on local ${\bf F}_0$ including the number of all higher genus curves.
By taking suitable limits $R \rightarrow 0$ and/or $\hbar \rightarrow 0$ we recover
four dimensional Seiberg-Witten theory or its coupling to graviphoton backgrounds.

\bigskip

\section{Adding matter and blow ups}
\setcounter{equation}{0}

According to the prescription of geometric engineering \cite{KKV,KMV,KV,CKYZ} matters in the 
fundamental
representation are obtained by blow ups. By making a blow up
at a point on the Hirzebruch surface ${\bf F}_0$ or ${\bf F}_1$,
we obtain the second del Pezzo surface $dP_2^{(0)}$. We also have its
\lq\lq cousin\rq\rq\ $dP_2^{(1)}$ from blowing up ${\bf F}_2$ or ${\bf F}_1$.
The reflexive polyhedra for these toric surfaces are 
No. 5 and 6, respectively, in the figure 1 of \cite{CKYZ}.
The dual toric diagrams of these surfaces are one loop pentagon diagrams 
and topological string amplitudes for $dP_2^{(m)}~(m=0,1)$ are
\beqa
Z_{top~str}^{(dP_2)}  &=& \sum_{R_1 \cdots R_5} W_{R_5 R_1} W_{R_1 R_2} \cdots W_{R_4 R_5}
\cdot e^{- \ell_{R_1} t_F - \ell_{R_2} t_B - \ell_{R_3} (t_F - t_E) -\ell_{R_4} t_E 
-\ell_{R_5} (t_B + (m+1)t_F - t_E)}\CR  
& &~\times (-1)^{\ell_{R_2} + \ell_{R_3} + \ell_{R_4}} 
\cdot (-1)^{m (\ell_{R_5}-\ell_{R_2})} 
\cdot q^{-\frac{1}{2} (\k_{R_2} + \k_{R_3} + \k_{R_4}) } 
\cdot q^{\frac{m}{2} (\k_{R_5} - \k_{R_2} ) }~.
\eeqa

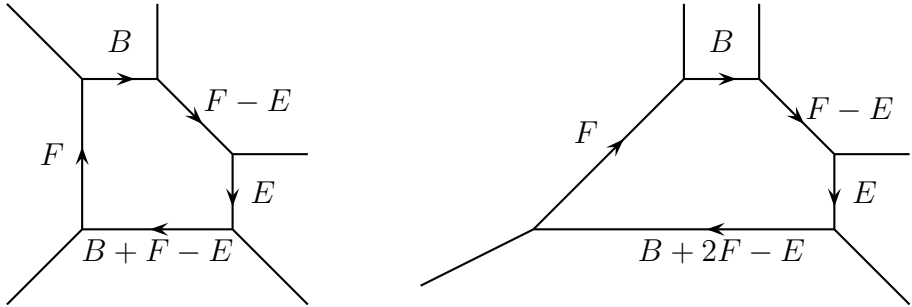
\begin{center}

\begin{pspicture}(-7,-2)(5,4) 

\psline[arrowsize=5pt]{->}(-6,0.5)(-6,1.6)
\psline(-6,1.5)(-6,2.5)
\psline[arrowsize=5pt]{->}(-6,2.5)(-5.3,2.5)
\psline(-5.4,2.5)(-5,2.5)
\psline[arrowsize=5pt]{->}(-5,2.5)(-4.4,1.9)
\psline(-4.5,2)(-4,1.5)
\psline[arrowsize=5pt]{->}(-4,1.5)(-4,0.8)
\psline(-4,0.9)(-4,0.5)
\psline[arrowsize=5pt]{->}(-4,0.5)(-5.1,0.5)
\psline(-5,0.5)(-6,0.5)
\psline(-6,0.5)(-7,-0.5)
\psline(-6,2.5)(-7,3.5)
\psline(-5,2.5)(-5,3.5)
\psline(-4,1.5)(-3,1.5)
\psline(-4,0.5)(-3,-0.5)
\rput(-6.4,1.5){$F$}
\rput(-3.8,2.2){$F-E$}
\rput(-3.6,1){$E$}
\rput(-5.5,3){$B$}
\rput(-5,0.2){$B+F-E$}


\psline[arrowsize=5pt]{->}(0,0.5)(1.2,1.7)
\psline(1.1,1.6)(2,2.5)
\psline[arrowsize=5pt]{->}(2,2.5)(2.7,2.5)
\psline(2.6,2.5)(3,2.5)
\psline[arrowsize=5pt]{->}(3,2.5)(3.6,1.9)
\psline(3.5,2)(4,1.5)
\psline[arrowsize=5pt]{->}(4,1.5)(4,0.8)
\psline(4,0.9)(4,0.5)
\psline[arrowsize=5pt]{->}(4,0.5)(2.3,0.5)
\psline(2.4,0.5)(0,0.5)
\psline(0,0.5)(-1.5,-0.25)
\psline(2,2.5)(2,3.5)
\psline(3,2.5)(3,3.5)
\psline(4,1.5)(5,1.5)
\psline(4,0.5)(5,-0.5)
\rput(0.7,1.8){$F$}
\rput(4.2,2.1){$F-E$}
\rput(4.4,1){$E$}
\rput(2.5,3){$B$}
\rput(2.5,0.2){$B+2F-E$}

\end{pspicture}
$B\cdot B = -(m+1)~,~F\cdot F =0~,~B\cdot F =+1$  \\
$E\cdot E = -1~,~E\cdot F = E\cdot B =0$ \\

Figure 3 : Toric diagram of the second del Pezzo surface 
${dP}_2^{(0)}$ (left) and ${dP}_2^{(1)}$ (right)

\end{center}


It is convenient to introduce the following building blocks
\beqa
K_{R_1 R_2}(Q) &:=& \sum_S Q^{\ell_S} W_{R_1S}(q) W_{SR_2}(q)~, \\
\hskip-7mm L_{R_1 R_2}(Q_1, Q_2) &:=& \sum_{S_1, S_2} Q_1^{\ell_{S_1}} Q_2^{\ell_{S_2}}
W_{R_1S_1}(q) W_{S_1S_2}(q)W_{S_2R_2}(q) 
(-1)^{\ell_{S_1} + \ell_{S_2}} q^{-\frac{1}{2} (\k_{S_1} + \k_{S_2})}.
\eeqa
Then by cutting the diagram at two internal lines with 
the K\"ahler modulus $t_B$, we can express the amplitude as
\beqa
Z_{top~str}^{(dP_2)}  &=& \sum_{R_1, R_2} Q_B^{\ell_{R_1} + \ell_{R_2}} Q_F^{(m+1)\ell_{R_2}} 
Q_E^{-\ell_{R_2}} (-1)^{(1-m)\ell_{R_1} + m \ell_{R_2}}
q^{-\frac{1}{2} ((1-m)\k_{R_1} + m\k_{R_2})} \CR  & &
~~~~~\times K_{R_1 R_2}(Q_F) \cdot L_{R_1 R_2} (Q_F Q_E^{-1}, Q_E)~,
\label{dP2a}\eeqa
where $Q_F = e^{-t_F}, Q_B = e^{-t_B}$ and $Q_E = e^{-t_E}$.


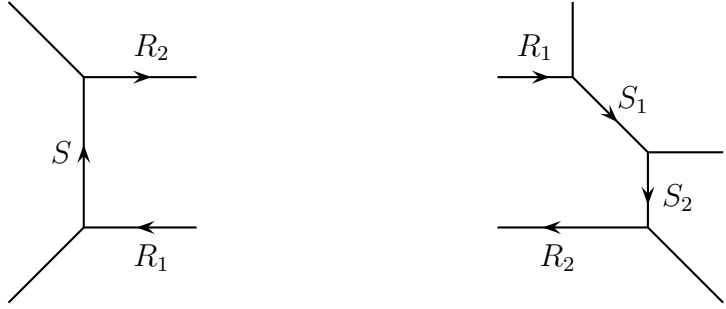
\begin{center}

\begin{pspicture}(-7,0)(5,4) 

\psline[arrowsize=5pt]{->}(-5.5,1)(-5.5,2.1)
\psline(-5.5,2)(-5.5,3)
\psline[arrowsize=5pt]{->}(-5.5,3)(-4.6,3)
\psline(-4.7,3)(-4,3)
\psline[arrowsize=5pt]{->}(-4,1)(-4.8,1)
\psline(-4.7,1)(-5.5,1)
\psline(-5.5,1)(-6.5,0)
\psline(-5.5,3)(-6.5,4)
\rput(-5.8,2){$S$}
\rput(-4.6,0.6){$R_1$}
\rput(-4.6,3.4){$R_2$}

\psline[arrowsize=5pt]{->}(0,3)(0.7,3)
\psline(0.6,3)(1,3)
\psline[arrowsize=5pt]{->}(1,3)(1.6,2.4)
\psline(1.5,2.5)(2,2)
\psline[arrowsize=5pt]{->}(2,2)(2,1.3)
\psline(2,1.4)(2,1)
\psline[arrowsize=5pt]{->}(2,1)(0.6,1)
\psline(0.7,1)(0,1)
\psline(1,3)(1,4)
\psline(2,2)(3,2)
\psline(2,1)(3,0)
\rput(1.8,2.7){$S_1$}
\rput(2.4,1.4){$S_2$}
\rput(0.5,3.4){$R_1$}
\rput(0.8,0.6){$R_2$}

\end{pspicture}

\begin{enumerate}
\item
The self-intersection of the internal two-cycle of $K_{R_1 R_2}(Q)$ is zero. 
\item
Two internal  two-cycles of $L_{R_1 R_2} (Q_1, Q_2)$ are $(-1)$ curves 
intersecting each other with intersection number $+1$. 
\end{enumerate}

Figure 4 : Building blocks $K_{R_1 R_2}(Q)$ (left) and $L_{R_1 R_2} (Q_1, Q_2)$ (right) 

\end{center}


As we have seen in the last section, we have an identity 
\beq
K_{R_1 R_2}(Q) =W_{R_1}(q)W_{R_2}(q)
\exp \left( \sum_{n=1}^\infty \frac{\widetilde{f}_{R_1R_2}(q^n)}{n}Q^n \right)~. \label{K-identity}
\label{KR}\eeq
Let us introduce a similar ansatz for $L_{R_1 R_2} (Q_1, Q_2)$;\\
{\bf Proposition 3}
\beqa
L_{R_1 R_2} (Q_1, Q_2) &=& W_{R_1}(q) W_{R_2}(q) \CR
& & \hskip-15mm \times\exp \left( \sum_{n=1}^\infty \frac{A_{R_1}(q^n)}{n} Q_1^n 
+ \sum_{n=1}^\infty \frac{A_{R_2}(q^n)}{n} Q_2^n
+ \sum_{n=1}^\infty \frac{A_{R_1 R_2}(q^n)}{n} (Q_1 Q_2)^n\right).
\label{L-ansatz}\eeqa
By comparing coefficients of $Q_1, Q_2$ and $Q_1Q_2$ of both sides of the above formula, 
we find
\beqa
(-1) W_{R_1\unitbox}(q) W_{\unitbox\! \,\,\bullet}(q) W_{\bullet R_2}(q)
&=& W_{R_1}(q) W_{R_2}(q) A_{R_1}(q)~, \\
 (-1) W_{R_1 \bullet}(q) W_{\bullet\! \,\,\unitbox}(q) W_{\unitbox R_2}(q)
&=& W_{R_1}(q) W_{R_2}(q) A_{R_2}(q)~, \\     
W_{R_1\unitbox}(q) W_{\unitbox\! \,\,\unitbox}(q) W_{\unitbox R_2}(q)
&=& W_{R_1}(q) W_{R_2}(q) 
\left( A_{R_1 R_2}(q) + A_{R_1}(q) A_{R_2}(q) \right)~. 
\eeqa
If one uses the relation
\beq
W_{R\! \,\,\unitbox} (q) = W_{\unitbox\! \,\,R} (q) 
= W_R(q) W_{\unitbox}^{-1} (q) \widetilde f_R(q)~,
\eeq
one can determine the unknown functions $A_{R_1},A_{R_2},A_{R_1R_2}$ as
\beq
A_{R_1}(q) =  - \widetilde f_{R_1} (q)~, \qquad
A_{R_2}(q) = -  \widetilde f_{R_2} (q)~,
\eeq
and 
\beqa
A_{R_1 R_2}(q) &=& \widetilde f_{R_1}(q) \widetilde f_{R_2}(q) 
\left( 1 +  W_{\unitbox}^{-2} (q)\right) - \widetilde f_{R_1}(q) \widetilde f_{R_2}(q)  \CR
&=& \widetilde f_{R_1 R_2}(q)~.
\eeqa
We present a proof of the above formula (\ref{L-ansatz}) making use of
the skew Schur functions \cite{Mac,ORV} in Appendix C.

Recall that the coefficients $C_k(R_1, R_2)$ were 
introduced in (\ref{Ck}) as the expansion coefficients of $f_{R_1R_2}(q)$. 
Similarly we define 
$C_k(R)$ as the expansion coefficients of
$f_{R}(q)$,
\beq
f_R(q)=\sum_k C_k(R)q^k.
\eeq
Note that $C_k(R)= C_k(R, \bullet)$.
Using these coefficients, we obtain the following expression for $L_{R_1 R_2}(Q_1, Q_2)$;
\beqa
L_{R_1 R_2} (Q_1, Q_2) &=& W_{R_1}(q) W_{R_2}(q) 
\exp \left\{ \sum_{n=1}^\infty \frac{1}{n} W_{\unitbox}^2 (q^n) 
\left( (Q_1Q_2)^n - Q_1^n  - Q_2^n  \right) \right\} \CR
& &~~~\times \prod_k  \frac{\left( 1 - q^k Q_1 \right)^{C_k(R_1)}
\left( 1 - q^k Q_2 \right)^{C_k(R_2)}}
{\left( 1 - q^k Q_1Q_2 \right)^{C_k(R_1, R_2)}}~. 
\label{LR}
\eeqa
Substituting (\ref{KR}) and (\ref{LR}) into (\ref{dP2a}), one finds
\beqa
Z_{top~str}^{(dP_2)} &=&\exp \left\{ \sum_{n=1}^\infty \frac{1}{n} 
W_{\unitbox}^2 (q^n) \left( 2 Q_F^n - (Q_F Q_E^{-1})^n  - Q_E^n  \right) \right\} \CR
& &~~\times\sum_{R_1, R_2} Q_B^{\ell_{R_1} + \ell_{R_2}} Q_F^{(m+1)\ell_{R_2}} 
Q_E^{-\ell_{R_2}} (-1)^{(1-m)\ell_{R_1} + m \ell_{R_2}}
q^{-\frac{1}{2} ((1-m)\k_{R_1} + m\k_{R_2})} \CR
& &~~\times W_{R_1}^2(q) W_{R_2}^2(q) 
\prod_k \frac{\left( 1 - q^k (Q_F Q_E^{-1}) \right)^{C_k(R_1)}
\left( 1 - q^k Q_E \right)^{C_k(R_2)}} {\left( 1 - q^k Q_F\right)^{2 C_k(R_1, R_2)}}~. \label{dP2}
\eeqa

\bigskip

\subsection{Perturbative one loop effective coupling}

In this topological string amplitude, the power of $Q_B$ is identified with
the space-time instanton number of Yang-Mills field. The first factor of (\ref{dP2}) 
is interpreted as the perturbative part of the free energy (the Seiberg-Witten
prepotential);
\beq
{\cal F}_{one~loop} = \sum_{n=1}^\infty \frac{1}{n} 
W_{\unitbox}^2 (q^n) \left( 2 Q_F^n - (Q_F Q_E^{-1})^n  - Q_E^n  \right)~.
\eeq
We make the following identification of parameters in the gauge theory side;
\beq
Q_F = e^{-4Ra}~, \quad Q_E = e^{-2R(a - m)} \quad {\rm and} 
~~~Q_F Q_E^{-1}  = e^{-2R(a + m)}~, \label{identify}
\eeq
where $a$ is vacuum expectation value of the scalar in the ${\cal N}=2$
vector multiplet and $m$ is the mass of the fundamental matter.
$R$ is the radius of the circle. The parameter $\hbar$ is the genus
expansion parameter in the topological string and we have
\beq
q = e^{-2\hbar R}~.
\eeq
When we take the limit $q \to 0$, 
\beq
W_{\unitbox}^2 (q^n) = \frac{1}{\left(e^{n\hbar R} - e^{-n\hbar R}\right)^2} 
\to (2n\hbar R)^{-2}~,
\eeq
and hence
\beq
{\cal F}_{one~loop} \rightarrow \frac{1}{4R^2} \sum_{n=1}^\infty \frac{1}{n^3} 
\left( 2 e^{-4nRa} - e^{-2nR(a+m)}  - e^{-2nR(a-m)}  \right)~.
\eeq
The one loop effective coupling is the second derivative of the prepotential;
\beq
\tau_{one~loop} = \frac{\partial^2 {\cal F}_{one~loop}}{ \partial a^2}  
= -8 \log (1-  e^{-4nRa}) + \log (1 - e^{-2nR(a+m)})( 1- e^{-2nR(a-m)})~.
\eeq
Up to a linear term in $a$ which should be provided from 
the classical part of the Yukawa couplings
we obtain
\beq
\tau_{one~loop} = -8 \log \sinh 2Ra + 
\log \sinh R(a+m)+ \log \sinh R(a-m)~,
\eeq
which has the correct five-dimensional form of summing up Kaluza-Klein modes. 
In the decompactified limit 
$R\rightarrow \infty$ one finds
\beq
\tilde{\tau}={\tau \over R}=-16|a|+|a+m|+|a-m|~.
\eeq
This gives the well-known behavior of the coupling constant of uncompactified five-dimensional 
gauge theory \cite{Seib}.

\bigskip

\subsection{Comparison with Nekrasov's conjecture}

Let us now compare our topological string amplitude with Nekrasov' formula
for gauge theory with matter. The instanton part of the amplitude
on local del Pezzo surface $dP_2$ is
\beqa
Z_{top~str}^{(dP_2)} &=&\sum_{R_1, R_2} Q_B^{\ell_{R_1} + \ell_{R_2}} Q_F^{2 \ell_{R_2}} 
Q_E^{-\ell_{R_2}} (-1)^{\ell_{R_2}} q^{-\frac{1}{2}\k_{R_2}} W_{R_1}^2(q) W_{R_2}^2(q) \CR
&\times&\prod_k \frac{\left( 1 - q^k (Q_F Q_E^{-1}) \right)^{C_k(R_1)}
\left( 1 - q^k Q_E \right)^{C_k(R_2)}} { \left( 1 - q^k Q_F\right)^{2 C_k(R_1, R_2)}}~,
\eeqa
where we chose $m=1$ for convenience.  As in the pure Yang-Mills case, we first replace
the summation over $R_2$ with that over $R_2^t$. Then the same computation
as before gives
\beqa
Z_{top~str}^{(dP_2)}&=&\sum_{R_1,R_2} \left( \frac{Q_B}{16 Q_F} \right)^{\ell_{R_1} + \ell_{R_2}} 
Q_F^{2 \ell_{R_2}} Q_E^{-\ell_{R_2}} (-1)^{\ell_{R_2}} q^{+\frac{1}{2}\k_{R_2}} \CR
& & \times \prod_k \left( 1 - q^k (Q_F Q_E^{-1}) \right)^{+C_k(R_1)}
\left( 1 - q^k Q_E \right)^{+C_k(R_2^t)}  \CR
& & \times \prod_{\ell, m \in \{1,2\}} \prod_{i,j = 1}^\infty
\frac{\sinh R \left( a_{\ell m} + \hbar (\m_{\ell,i} - \m_{m,j} + j -i) \right)}
{\sinh R \left( a_{\ell m} + \hbar ( j -i) \right)}~.
\eeqa
Using the relation $C_k(R^t) =  C_{-k}(R)$, we have
\beqa
& &\prod_k \left( 1 - q^k (Q_F Q_E^{-1}) \right)^{+C_k(R_1)}
\left( 1 - q^k Q_E \right)^{+C_k(R_2^t)}  \CR
& &~= 2^{\ell_{R_1} + \ell_{R_2}} q^{\frac{1}{4}(\k_{R_1} - \k_{R_2})} 
Q_F^{\frac{1}{2}\ell_{R_1}}  Q_E^{\frac{1}{2}(\ell_{R_2} - \ell_{R_1})}  \CR
& & \times \prod_k \left( \sinh R(a+m + \hbar k) \right)^{C_k(R_1)}
\left( \sinh R(a-m - \hbar k) \right)^{C_k(R_2)}~.
\eeqa
Hence we obtain
\beqa
Z_{top~str}^{(dP_2)}&=&\sum_{R_1,R_2} \left( \frac{Q_B}{8 Q_F} \right)^{\ell_{R_1} + \ell_{R_2}} 
Q_F^{2 \ell_{R_2}+\frac{1}{2}\ell_{R_1}}  Q_E^{-\frac{1}{2}(\ell_{R_1} + \ell_{R_2})} 
 q^{\frac{1}{4}(\k_{R_1} + \k_{R_2})}  \CR
& & \times \prod_k \left( \sinh R(a+m + \hbar k) \right)^{C_k(R_1)}
\left( \sinh R(-a+m +\hbar k) \right)^{C_k(R_2)} \CR
& & \times \prod_{\ell, m \in \{1,2\}} \prod_{i,j = 1}^\infty
\frac{\sinh R \left( a_{\ell m} + \hbar (\m_{\ell,i} - \m_{m,j} + j -i) \right)}
{\sinh R \left( a_{\ell m} + \hbar ( j -i) \right)}~.
\eeqa
In the four dimensional limit $R \to 0$ last two factors give
\beqa
& &\frac{1}{R^{3(\ell_{R_1}+\ell_{R_2})}}
\prod_k \left( a+m + \hbar k \right)^{C_k(R_1)}
\left( -a+m +\hbar k \right)^{C_k(R_2)} \CR
& &~\times \prod_{\ell, m \in \{1,2\}} \prod_{i,j = 1}^\infty
\frac{  a_{\ell m} + \hbar (\m_{\ell,i} - \m_{m,j} + j -i) }
{  a_{\ell m} + \hbar ( j -i) }~. \label{4Dmatter}
\eeqa
We note that $Q_B = 2 ( R \Lambda)^3$ and the powers of $R$ cancel
each other.

On the other hand Nekrasov's formula for $SU(2)$ theory with $N_f=1$ is given by
(in our notation);
\beqa
& &\left( \hbar \Lambda \right)^{\ell_{R_1} + \ell_{R_2}} 
\prod_{(\ell, i)} \frac{\Gamma(\frac{a_\ell +m}{\hbar} +1 + \m_{\ell, i} -i)}
{\Gamma(\frac{a_\ell +m}{\hbar} + 1 -i)} \CR
& &~~\times \prod_{\ell, m \in \{1,2\}} \prod_{i,j = 1}^\infty
\frac{a_{\ell m} + \hbar (\m_{\ell,i} - \m_{m,j} + j -i)}
{a_{\ell m} + \hbar ( j -i)}~.
\label{dPcompu}\eeqa
Since we have seen that the pure gauge part agrees in section 3, we will examine 
the additional factor of matter contribution (first line of (\ref{dPcompu}));
\beq
\prod_{i=1}^{d(\m^{R_1})} \prod_{j=1}^{\m_{1,i}} \left( a + m + \hbar (j-i) \right)
\times \prod_{i=1}^{d(\m^{R_2})} \prod_{j=1}^{\m_{2,i}} \left( - a + m + \hbar (j-i) \right)~,
\eeq
where we have used $\Gamma(z+1) = z \Gamma(z)$ and absorbed the powers of $\hbar$. 
If we recall that
the generation function of $C_k(R)$ is
\beq
f_R(q) = \sum_{i=1}^{d(\m^R)} \sum_{j=1}^{\m_i} q^{j-i}~,
\label{fRq}\eeq
we see that the four dimensional limit of our amplitude in fact exactly
agrees with Nekrasov's formula.

In this comparison we have considered the amplitude for $dP_2^{(1)}$.
In the four dimensional limit the amplitudes for $dP_2^{(0)}$ and $dP_2^{(1)}$ are
simply related  by the sign flip of the mass parameter $m \to -m$.

\bigskip

\subsection{ Four dimensional limit and Seiberg-Witten prepotential}

Let us next check if our amplitude 
reproduces the Seiberg-Witten prepotential of $SU(2)$ theory
with a single hypermultiplet.

We first recall the definition of instanton expansion of the free-energy
\beqa
&& \log Z_{top\hskip1mm str}(Q_B,Q_F,Q_E,q)= {\cal F}_{one \hskip1mmloop}(Q_F,Q_E,q)+
{\cal F}_{inst}(Q_B,Q_F,Q_E,q)~,\\
&&{\cal F}_{inst}(Q_B,Q_F,Q_E,q)=\sum_{k=1}^{\infty} \sum_{n=1}^{\infty}{1\over n}
Q_B^{\,\,nk}{\cal F}_k(Q_F^{\,\,n},Q_E^{\,\,n},q^n)~,
\\
&&{\cal F}_k(Q_F,Q_E,q)=\sum_{g=0}^{\infty}{1\over (2\sin \frac{g_s}{2})^{2-2g}}\,f_g^{(k)}(Q_F,Q_E)~,
\label{dPinst}\eeqa
where $f_g^{(k)}$ gives the genus-$g$ $k$-instanton amplitudes and the string coupling constant $g_s$ is related to $q$ as
\beq
q=e^{ig_s}=e^{-2\hbar R}.
\eeq
Note that (\ref{dPinst}) defines the genus expansion of the Gopakumar-Vafa type. When we take the
four-dimensional limit, it is converted into an ordinary genus expansion in powers of $\hbar$. 

By explicit computations of (\ref{dP2}) we find the following results for genus zero instanton 
amplitudes
\beqa
f_{g=0}^{(1)}(Q_F,Q_E)&=&{N_0^{(1)}\over (Q_F-1)^2}~,\\ 
f_{g=0}^{(2)}(Q_F, Q_E) &=& \frac{N_0^{(2)} }{(Q_F+1)^2 (Q_F-1)^6}~,   \\
f_{g=0}^{(3)}(Q_F, Q_E)  &=& \frac{N_0^{(3)}}{(Q_F^2 + Q_F +1)^2 (Q_F -1)^{10} }~,
\eeqa
where the numerators are 
\beqa
N_0^{(1)} &=&  -(Q_F+1) + 2 Q_F Q_E^{-1}~,  \\
N_0^{(2)} &=&  (6Q_F^2 + 8Q_F +6)Q_F^2 
- ( 5Q_F^3 + 15 Q_F^2 + 15 Q_F +5) Q_F^2 Q_E^{-1}  \CR
& &~~+  (6Q_F^2 + 8Q_F +6) Q_F^3  Q_E^{-2}~,  \\
N_0^{(3)} &=& -(27 Q_F^5 + 70 Q_F^4 + 119 Q_F^3 + 119 Q_F^2
+ 70 Q_F + 27) Q_F^3  \CR
& &~~+ (32 Q_F^6 + 144 Q_F^5 + 288 Q_F^4 + 368 Q_F^3 + 288 Q_F^2
+ 144 Q_F +32) Q_F^3 Q_E^{-1}  \CR
& &~~- (7 Q_F^7 + 79 Q_F^6 + 216 Q_F^5 + 346 Q_F^4
+ 346 Q_F^3 + 216 Q_F^2 + 79 Q_F + 7) Q_F^3 Q_E^{-2}  \CR
& &~~+ ( 8 Q_F^7 + 46 Q_F^6+ 100 Q_F^5 + 124 Q_F^4 
+ 100 Q_F^3 + 46 Q_F^2 +8 Q_F) Q_F^3Q_E^{-3} ~.
\eeqa
The coefficients of the Taylor expansion of $f_{g=0}^{(k)}$ give
the integer invariants of Gopakumar-Vafa. We find a perfect
agreement with Table 3 in \cite{CKYZ}.
(Their K\"ahler moduli $t_i$ are related to ours by
$t_1= t_F, t_2=t_B, t_3=t_E + t_F$.)
We note that when the mass $m$ goes to infinity, $Q_E \to 0$ and
the most singular part in this limit reproduces the result of the
Hirzebruch surface (we redefine $Q_BQ_F/Q_E$ as $Q_B$ for the Hirzebruch surface). 

Substituting (\ref{identify}), we find
\beqa
&& f_{g=0}^{(1)}(Q_F, Q_E) \approx \frac{1}{4} \frac{m}{a^2} R^{-1}, \quad
f_{g=0}^{(2)}(Q_F, Q_E) \approx \frac{1}{2^{10}} \frac{5m^2 - 3a^2}{a^6} R^{-4} ,   \CR
&& f_{g=0}^{(3)}(Q_F, Q_E)  \approx \frac{1}{3 \cdot 2^{13} } \frac{m(9m^2 - 7a^2)}{a^{10}} R^{-7} .
\eeqa
At genus zero we have a contribution $(2\hbar R)^{-2}$ from $\sin^{-2} (q/2)$ and
we identify
\beq
Q_B = 2 (R\Lambda)^3~,
\eeq
then in the four-dimensional limit $R \to 0$ we have
\beqa
&&f_{g=0}^{(1)}(Q_F, Q_E)\cdot Q_B \to \frac{1}{2^3} \frac{m}{a^2} \Lambda^3,\quad
f_{g=0}^{(2)}(Q_F, Q_E) \cdot Q_B^2 \to \frac{1}{2^{10}} \frac{5m^2 - 3a^2}{a^6} \Lambda^6,   \CR
&&f_{g=0}^{(3)}(Q_F, Q_E) \cdot Q_B^3 \to \frac{1}{3 \cdot 2^{12} } \frac{m(9m^2 - 7a^2)}{a^{10}} \Lambda^9.
\eeqa
The first two expressions agree with Eq.(5.5) of \cite{DKP}.
After a rescaling
\beq
m \to \frac{m}{\sqrt 2}~, \hskip 1cm  \Lambda \to \frac{\Lambda}{\sqrt 2}~, 
\eeq
we obtain
\beq
{\cal F}_1 = \frac{1}{2^5} \frac{m}{a^2} \Lambda^3, \quad
{\cal F}_2 = \left(  \frac{5}{2^{14}} \frac{m^2}{a^6}  
-  \frac{3}{2^{13}} \frac{1}{a^4} \right) \Lambda^6,   \quad
{\cal F}_3 = \left( \frac{3}{2^{18} } \frac{m^3}{a^{10}} 
- \frac{7}{3 \cdot 2^{17} } \frac{1}{a^8} \right)\Lambda^9.
\eeq
Then we find an agreement with \cite{Ohta} (Addendix D)
(up to an overall normalization factor of 2).
The decoupling limit $m\Lambda^3 \equiv \Lambda_0^4$ and $m \to \infty$ gives
the instanton expansion of the prepotential of pure $SU(2)$ gauge theory.

\bigskip

\bigskip

\subsection{Genus one amplitude and coupling to gravity}

In type II string theory
topological string amplitudes at higher genus 
compute the (F-type) couplings of four-dimensional gauge theory with 
gravitational background field \cite{BCOV}. In particular the genus one
amplitude ${\cal F}^{(1)}$ contributes to the following coupling;
\beq
\frac{1}{2}{\cal F}^{(1)} (a,m) \left( \chi - \frac{3}{2}\sigma \right)~,
\eeq
where $\chi$ and $\sigma$ are the Euler characteristic and the 
signature of the four manifold.

Again by explicit calculations of (\ref{dP2}) we can derive 
that the $k$-instanton parts of the free 
energy $f_{g=1}^{(k)}(Q_F, Q_E) $ at genus one. They are given by
\beqa
f_{g=1}^{(1)}(Q_F, Q_E) &=& 0~,  \\
f_{g=1}^{(2)}(Q_F, Q_E) &=& \frac{N_1^{(2)} }{(Q_F+1)^2 (Q_F-1)^8}~,   \\
f_{g=1}^{(3)}(Q_F, Q_E)  &=& \frac{N_1^{(3)}}{(Q_F^2 + Q_F +1)^2 (Q_F -1)^{12} }~,
\eeqa
where the numerators are 
\beqa
N_1^{(2)} &=&  (9Q_F^2 + 14Q_F +9)Q_F^3 
- ( 8Q_F^3 + 24 Q_F^2 + 24 Q_F +8)Q_F^3 Q_E^{-1}  \CR
& &~~+  (9Q_F^2 + 14Q_F +9) Q_F^4  Q_E^{-2}~,  \\
N_1^{(3)} &=& -(10 Q_F^7 + 188 Q_F^6 + 508 Q_F^5+ 
830 Q_F^4 + 830 Q_F^3 + 508 Q_F^2
+ 188 Q_F + 10) Q_F^3  \CR 
& &~\hskip-5mm+ (9 Q_F^8 + 254 Q_F^7 + 1033 Q_F^6 + 2030 Q_F^5 
+ 2564 Q_F^4 + 2030 Q_F^3 + 1033 Q_F^2
+ 254 Q_F +9) Q_F^3 Q_E^{-1}  \CR
& &~\hskip-5mm - (72 Q_F^8 + 584 Q_F^7 + 1536 Q_F^6 + 2416 Q_F^5 + 2416 Q_F^4
+ 1536 Q_F^3 + 584 Q_F^2 + 72 Q_F ) Q_F^3 Q_E^{-2}  \CR
& &~\hskip-5mm + ( 68 Q_F^8 + 336 Q_F^7+ 692 Q_F^4 + 880 Q_F^5 
+ 692 Q_F^4 + 336 Q_F^3 + 68 Q_F^2) Q_F^3Q_E^{-3} ~.
\eeqa
The coefficients of the Taylor expansion of $f_{g=1}^{(k)}(Q_F, Q_E) $ give
the genus-one integer invariants of Gopakumar-Vafa.  We note that
there is no contribution at genus one from one-instanton sector.
In the decoupling limit; $Q_E \to 0$, the most singular term
reproduces the genus one amplitudes of the Hirzebruch surface.
To take the four dimensional limit, we substitute 
$Q_F = \exp(-4Ra),~Q_E = \exp(-2R(a-m))$ 
and $Q_B = 2 (R\Lambda)^3$ as in the genus zero computation. 
Then we have
\beqa
f_{g=1}^{(2)}(Q_F, Q_E) \cdot Q_B^2 
&\rightarrow& \frac{1}{2^{11}} \frac{4m^2 - 3a^2}{a^8} \Lambda^{6}, \label{genus1a}  \\
f_{g=1}^{(3)}(Q_F, Q_E)  \cdot Q_B^3
&\rightarrow& \frac{1}{3 \cdot 2^{13} } 
\frac{m(32m^2 - 27a^2)}{a^{12}} \Lambda^{9}~.
\label{genus1b}\eeqa

We can compare the above result with the gravitational coupling in the 
twisted ${\cal N}=2$ $SU(2)$ Yang-Mills theory with a fundamental matter.
For pure Yang-Mills theory such comparison has been made in \cite{KMT,DST}.
The anomaly cancellation of the measure of the $u$-plane integral
of the Seiberg-Witten theory predicts the following coupling to the gravity \cite{MW};
\beq
\frac{\chi}{2} \log \left( \frac{1}{\Lambda} \frac{du}{da} \right)
+ \frac{\sigma}{8} \log \left( \frac{\Delta}{\Lambda^6} \right)~.
\eeq
Here $u$ is the coordinate on the moduli space of the Coulomb branch,
$a(u)$ is the period of the Seiberg-Witten differential on the $\alpha$ cycle
and $\Delta$ is the discriminant of the Seiberg-Witten curve.
As we know, the curve for $N_f =1$ is given by
\beq
y^2 = (x^2 - u)^2 - \Lambda^3 (x+m)~,
\eeq
with the discriminant
\beq
\Delta = 27 \Lambda^6 + 256 \Lambda^3 m^3 - 288 \Lambda^3 mu 
- 256 m^2 u^2 + 256 u^3~.
\eeq
The inversion of the period $a=a(u)$ in the weak coupling region is \cite{Ohta}
\beq
u(a) = 2 a^2 + \frac{m}{16 a^2} \Lambda^3 - \frac{3}{2048 a^4} \Lambda^6
+ \frac{5 m^2}{4096 a^6} \Lambda^6 - \frac{7 m}{65536 a^8} \Lambda^9
+ \frac{9 m^3}{131072 a^{10}} \Lambda^9 +\cdots~.
\eeq
The twisted theory and the physical theory agree on the hyperK\"ahler
manifold and we will take $K3$ surface with $\chi = 24$ and $\sigma = -16$.
Then comparing the gravitational coupling of both theories, we find
\beq
{\cal F}^{(1)} = \frac{1}{2} \log \left( \frac{1}{\Lambda} \frac{du}{da} \right)
- \frac{1}{12} \log \left( \frac{\Delta}{\Lambda^6} \right)~.
\eeq
Substituting the above data obtained from the Seiberg-Witten curve
we compute 
\beq
{\cal F}^{(1)} = \frac{1}{12} \log \frac{4a^2}{2a^2 -m^2} 
+ \frac{1}{2^{14}}\frac{2m^3 -3 a^2}{a^8} \Lambda^6 +
\frac{1}{3 \cdot 2^{18}} \frac{m(16 m^2 - 27 a^2)}{a^{12}} \Lambda^9 + \cdots~.
\eeq
After the same rescaling of $m$ and $\Lambda$ we have used in the genus zero,
we find a precise agreement with our prediction (\ref{genus1a}),(\ref{genus1b}).

\section{Discussions}

In this paper we have used geometric transition and computed all genus topological string 
amplitude on the local Calabi-Yau manifold $K_{{\bf F}_0}$ and have shown that the results agree 
with 
the Nekrasov's formula for ${\cal N}=2$ gauge theory with 
parameters $\beta$, $\hbar$ being kept non-zero.
When one takes the four-dimensional limit $\beta=R\rightarrow 0$, one recovers the well-known 
Seiberg-Witten theory and also its coupling to external graviphoton background.

We have also computed the amplitude for the local 2nd del Pezzo surface and have shown that the
result agrees with Nekrasov's formula for gauge theory with a matter hypermultiplet. 

We have provided proofs for formulas involving Hopf-link invariants which 
appeared in Chern-Simons calculations. 
In this sense our derivations are rigorous given the starting point of Chern-Simons
amplitudes.
We feel, however, that there is at present a lack of efficient
mathematical techniques in 
handling formulas containing Chern-Simons invariants and a need for developing more powerful 
machinery. One obvious candidate is the operator method of 
two-dimensional CFT which is now being developed in \cite{AKMV,ORV}. 

In a future publication we would like to study more details of topological string theory by 
making use of operator techniques.

\vskip10mm
\begin{center}
{\bf Acknowledgements}
\end{center}

H.K. would like to thank M. Mulase, H. Ohta and A. Tsuchiya for helpful discussions.
Our research is supported in part 
by the Grant-in-Aid for Scientific Research (No.15540253 and No. 14570073)
from Japan ministry of education, culture and sports.

\vskip10mm
\begin{center}
{\bf Notes added}
\end{center}

After this paper was submitted to e-print archives, a new paper has appeared \cite{HIV} which is
closely related to ours. 
The formula (2.4) has appeared in the mathematics literature \cite{Zhou},
where the degree of the normal bundle is used instead of the self-intersection number.
We would like to thank the referee for this comment.


\setcounter{section}{0}

\section*{Appendix A : Chern-Simons invariants}
\renewcommand{\theequation}{A.\arabic{equation}}\setcounter{equation}{0}

The primary fields of $U(N)$ WZW theory are associated with
highest weight representation at level $N+k$. The space of
conformal blocks of WZW theory on torus is the set of such
states $|R\rangle$. The generators $T$ and $S$ of $SL(2, {\bf Z})$ 
act on the space of conformal blocks. From the relation of
the space of conformal blocks and the Hilbert space of the
Chern-Simons theory on a 3-manifold  whose boundary is
a torus the matrix element $W_{R_1 R_2}
= \langle \bar R_1| S^{-1} |R_2\rangle$ is the Chern-Simons
invariant of the Hopf link in $S^3$ with linking number $+1$ \cite{Witten}.
It is expressed in terms of the $q$-numbers
\beq
[x] = q^{\frac{x}{2}} - q^{-\frac{x}{2}}~, \quad 
\eeq
where $q = \exp (\frac{2\pi i}{N+k})$. 
Let $\mu_j$ be the number of boxes in the $j$-th row of
the Young diagram $\mu^R$ associated with the representation $R$. 
We define two integers $\ell_R$ and $\kappa_R$ by \beq
\ell_R = \sum_{j=1}^{d(\mu^R)} \mu_j~, \quad 
\kappa_R = \ell_R + \sum_{j=1}^{d(\mu^R)} \mu_j (\mu_j -2j)~,
\eeq
where $d(\mu^R)$ is the number of rows of $\mu^R$.

Let $\hat\mu^R$ denote the transposed Young diagram of $\mu^R$ 
obtained by exchanging rows and columns. According to the Jacobi-Trudy formula
the Schur polynomial $s_{\mu^R}$ in the variables $(x_1, \cdots, x_N)$ corresponding 
to the Young diagram $\mu^R$ can be expressed in terms of 
the elementary symmetric polynomials $e_i(x_1, \cdots, x_N)$ as follows;
\beq
s_{\mu^R}= \det_{r \times r } M_{\mu^R}~,
\eeq
where $r = d(\hat\mu^R)$ and $M_{ij} = e_{\hat{\mu_i} +j -i}$. 
This formula is extended to the generalized Schur polynomial $s_{\mu^R}( E(t) )$ 
for any formal power series $E(t) =  1 + \sum_{n=1}^\infty a_n t^n$ by replacing
the generating function of the elementary symmetric polynomials
$P(t) = \prod_{i=1}^N (1 + x_i t) =   1 + \sum_{n=1}^\infty e_n t^n$ by $E(t)$.
Then the matrix element $W_{R_1, R_2}$ is given by \cite{ML}
\beq
W_{R_1, R_2} (q)  = (\dim_q R_1) \cdot q^{\ell_{R_2}/2} \cdot s_{\mu^{R_2}} ( E_{\mu^{R_1}}(t))~,
\label{Hopf}
\eeq
where 
\beq
 \dim_q R = \prod_{1 \leq i < j \leq d(\mu^R)} \frac{[\mu_i -\mu_j + j -i ]}{[j-i]}
\prod_{i=1}^{ d(\mu^R)} 
\prod_{k=1}^{\mu_i}{1\over [k-i + d(\mu^R)]}~,
\eeq
is called quantum dimension of $R$ and equal to the Chern-Simons invariant of the
unknot carrying representation $R$. The polynomial $E_{\mu^R}$ is
defined by
\beq
E_{\mu^R}(t) = \left( \prod_{j=1}^{d(\mu^R)} \frac{1+ q^{\mu_j - j} t}{1 + q^{-j} t} \right)
E_{\emptyset}(t)~, \hskip10mm
E_{\emptyset}(t)=1 + \sum_{n=1}^\infty ( \prod_{i=1}^n \frac { 1}{q^i -1}) t^n ~.
\label{Emu}\eeq
When $R_1$ is the trivial representation, (\ref{Hopf}) implies that the quantum dimension is
also expressed by the Schur functions;
\beq
W_{\bullet R}=W_R = \dim_q R =  q^{\ell_{R}/2}~s_{\mu^{R}} ( E_{\emptyset}(t))~.
\eeq

\bigskip


\section*{Appendix B : Proof of Proposition 2}
\renewcommand{\theequation}{B.\arabic{equation}}\setcounter{equation}{0}

We first recall the definition of $C_k (R_1, R_2)$;
\beq
\sum_k C_k(R_1, R_2) q^k = (q + q^{-1} -2) f_{R_1}(q) f_{R_2}(q) + f_{R_1}(q) + f_{R_2}(q)~,
\eeq
where $f_R(q)$ is given by (\ref{fRq}).
To simplify the notation let us use the following abbreviation;
\beq
d_1 = d(\m^{R_1})~, \quad d_2 = d(\m^{R_2})~, \quad \m_{1,i} = \m_i~, \quad \m_{2,j} = \n_j~,
\eeq
and set $R=1$ for simplicity.
Paying attention to the fact that the second argument is the transpose of $R_2$,
\beqa
\sum_{k} C_k( R_1, R_2^t) q^k  
&=& f_{R_1}(q)\sum_{j=1}^{d_2} \sum_{i=1}^{\n_j}  \left( q^{1+ j - i} + q^{-1 + j - i} -2 q^{j - i} \right) 
+ f_{R_1}(q) + \left( \sum_{j=1}^{d_2} \sum_{i=1}^{\n_j} q^{j - i} \right) \CR
&=& f_{R_1}(q) q^{d_2} + f_{R_1}(q)  \sum_{j=1}^{d_2} \left( q^{j - \n_j -1} -  q^{j - \n_j} \right) 
+ \left( \sum_{j=1}^{d_2} \sum_{i=1}^{\n_j} q^{j - i} \right)~.
\eeqa
Hence we obtain
\beqa
& & \prod_k \frac{1}{ \sinh (2a + \hbar k)^{C_k(R_1, R_2^t)}}  \CR
&=&~\prod_{i=1}^{d_1} \prod_{j=1}^{\m_i} \frac{1}{\sinh (2a + \hbar(j - i + d_2))} 
 \times \prod_{i=1}^{d_1} \prod_{j=1}^{d_2} \prod_{k=1}^{\m_i} 
 \frac{\sinh \left( 2a + \hbar (k - i +j - \n_j ) \right) }{\sinh \left (2a + \hbar(k - i +j -1 - \n_j  )\right) }  \CR
& &~~ \times \prod_{j=1}^{d_2} \prod_{i=1}^{\n_j}  \frac{1}{\sinh (2a + \hbar(j - i))}  \CR
&=& ~\prod_{i=1}^{d_1} \prod_{j=1}^{\m_i} \frac{1}{\sinh (2a + \hbar(j - i + d_2))}  
\times \prod_{i=1}^{d_1} \prod_{j=1}^{d_2}  \frac{\sinh \left( 2a + \hbar (\m_i - \n_j +j -i ) \right) }
{\sinh \left (2a + \hbar(j - i - \n_j)\right) } \CR
& &~~ \times \prod_{j=1}^{d_2} \prod_{i=1}^{\n_j}  \frac{1}{\sinh (2a + \hbar(j - i))}~.  \label{stringside}
\eeqa

On the other hand, splitting the set of indices $(i,j) \in {\bf Z}_+ \times {\bf Z}_+$
into four disjoint sets; $\{ 1 \leq i \leq d_1, 1 \leq j \leq d_2 \} \cup
\{ d_1 < i < \infty, 1 \leq j \leq d_2 \} \cup \{ 1 \leq i \leq d_1, d_2 < j < \infty \} \cup
\{ d_1 < i < \infty, d_2 < j < \infty \}$, we can compute 
the expression in the gauge theory side;
\beqa
& & \prod_{i,j=1}^\infty \frac{ \sinh \left( 2a + \hbar(\m_{i} - \n_{j} + j - i) \right)}
{\sinh \left( 2a + \hbar ( j- i) \right) } \CR
&=& \prod_{i=1}^{d_1} \prod_{j=1}^{d_2} \frac{ \sinh \left( 2a + \hbar(\m_{i} - \n_{j} + j - i) \right)}
{\sinh \left( 2a + \hbar ( j- i) \right) }
\times \prod_{j=1}^{d_2} \prod_{i=1}^{\n_j} \frac{1}{\sinh \left( 2a + \hbar ( j- i -d_1) \right) }  \CR
& &~~~~~~\times \prod_{i=1}^{d_1} \prod_{j=1}^{\m_i}\frac{1}{\sinh \left( 2a + \hbar ( j- i + d_2) \right) }~.
\eeqa
To compare with (\ref{stringside}) we can further transform the first two factors into 
\beqa
& &\prod_{i=1}^{d_1} \prod_{j=1}^{d_2} \sinh \left( 2a + \hbar(\m_{i} - \n_{j} + j - i) \right) 
\times \prod_{j=1}^{d_2} \prod_{i=1}^{d_1 + \n_j} \frac{1}{\sinh \left( 2a + \hbar ( j- i) \right) } \CR
&=&~\prod_{i=1}^{d_1} \prod_{j=1}^{d_2} \frac{\sinh \left( 2a + \hbar(\m_{i} - \n_{j} + j - i) \right)}
{\sinh \left(2a + \hbar(j - i - \n_j ) \right)} 
\times \prod_{j=1}^{d_2} \prod_{i=1}^{\n_j} \frac{1}{\sinh \left( 2a + \hbar ( j- i) \right) }~. 
\eeqa
Thus we find that the final expression is exactly the same as that of topological string side 
(\ref{stringside})!

\bigskip


\section*{Appendix C : Proof of Proposition 3}
\renewcommand{\theequation}{C.\arabic{equation}}\setcounter{equation}{0}
\renewcommand{\thesubsection}{C.\arabic{subsection}}\setcounter{subsection}{0}

We recall that the Hopf link invariants are given by 
\beq
W_{R_1, R_2} (q) = W_{R_1}(q) q^{{\ell_{R_2}\over 2}} s_{\m^{R_2}} \left( E_{\m^{R_1}} \right)
= W_{R_1}(q) s_{R_2} \left( q^{\m^{R_1} + \rho} \right)~,
\eeq
where $q^{\m + \rho}$ means that we make the following substitution;
\beq
s_R (x_i = q^{\m_i - i + \frac{1}{2}})~.
\eeq
Here we have made a shift of $1/2$ of the power of $q$ from (\ref{specialize}) 
in order to absorb the factor $q^{\ell_{R_2}/2}$.
By taking $R_1= \bullet$, we obtain
\beq
W_R(q) = s_R(q^\rho) = q^{\k_R/2} s_{R^t}(q^\rho)~, \label{dimension}
\eeq
and hence
\beq
W_{R_1 R_2}(q) = s_{R_1}(q^\rho) s_{R_2} (q^{\m^{R_1} + \rho})~. \label{link}
\eeq
According to \cite{AKMV} the topological vertex is given by
\beq
C_{R_1, R_2, R_3} = q^{\k_{R_2}/2+ \k_{R_3}/2} 
\sum_{Q_1, Q_2} N_{Q_1 Q_2}^{R_1 R_3^t} \frac{W_{R_2^t Q_1} W_{R_2 Q_2}}{W_{R_2}}~,
\eeq
where
\beq
N_{Q_1 Q_2}^{R_1 R_3^t} = \sum_Q N_{Q Q_1}^{R_1} N_{Q Q_2}^{R_3^t}~.
\eeq
Recall the definition of the skew Schur function \cite{Mac}
\beq
s_{R_1/R} (x) = \sum_{R_2} N_{R R_2}^{R_1} s_{R_2}(x)~.
\eeq
Hence we have
\beqa
C_{R_1, R_2, R_3} 
&=& q^{\k_{R_2}/2+ \k_{R_3}/2} 
\sum_{Q_1, Q_2, Q_3} N_{Q_3 Q_1}^{R_1} N_{Q_3 Q_2}^{R_3^t}~
 s_{R_2^t}(q^\rho) s_{Q_1}(q^{\m^{R_2^t}+\rho})
s_{Q_2}( q^{\m^{R_2} + \rho)}~, \CR
&=& q^{\k_{R_2}/2+ \k_{R_3}/2}  \sum_{Q_3} 
s_{R_2^t}(q^\rho) s_{R_1/Q_3}(q^{\m^{R_2^t} + \rho}) s_{R_3^t/ Q_3} ( q^{\m^{R_2} + \rho})~.
\eeqa
This is slightly different from the expression given in \cite{ORV}, 
but this form is more convenient in the following.
By taking $R_2 = \bullet$ and using the cyclic symmetry of $C_{R_1, R_2, R_3}$, we obtain
\beq
C_{\bullet, R_3, R_1} = q^{\k_{R_3}/2} \sum_{Q} 
s_{R_1/Q}(q^{\rho}) s_{R_3^t/ Q} ( q^{\rho})~.
\eeq
Hence,
\beq
W_{R_1 R_2} 
= q^{\k_{R_2}/2} C_{\bullet R_1 R_2^t}= q^{\k_{R_1}/2 + \k_{R_2}/2}
\sum_Q s_{R_2^t/Q}(q^{\rho}) s_{R_1^t/ Q} ( q^{\rho})~. \label{symmetric}
\eeq
Thus we have obtained a manifestly symmetric form of $W_{R_1 R_2}$.
On the other hand, if we take $R_3=\bullet$, only the trivial representation
contributes for the summation over $Q$ and
\beq
C_{R_1, R_2, \bullet} = s_{R_2}(q^{\rho}) s_{R_1}(q^{\m^{R_2^t} + \rho})~.
\eeq
Hence,
\beq
W_{R_1 R_2} = q^{\k_{R_2}/2} s_{R_2^t}(q^{\rho}) s_{R_1}(q^{\m^{R_2}+ \rho})
                     = s_{R_2}(q^{\rho}) s_{R_1}(q^{\m^{R_2}+ \rho})~.
\eeq
After the exchange of $R_1$ and $R_2$ we recover  our original expression of $W_{R_1 R_2}$.

When the topological vertices are expressed in terms of the (skew) Schur functions,
a summation over representations may be performed by using the following formulas \cite{Mac};
\beqa
\sum_R s_{R^t}(x) s_R(y) &=& \prod_{i,j \geq 1} (1 + x_i y_j)~, \label{schur1}\\
\sum_R s_{R}(x) s_R(y) &=& \prod_{i,j \geq 1} (1 - x_i y_j)^{-1}~. \label{schur2}
\eeqa

Next we introduce the \lq\lq relative\rq\rq\ hook length by
\beq
h_{R_1 R_2}(i,j) := \m_i^{R_1} -i  +\m_j^{R_2} - j +1~.
\eeq
When $R_1=R, R_2 = R^t$ it reduces to the standard hook length.
We have the following lemma;

{\bf Lemma}
$$
\prod_{i,j \geq 1} \left( 1 - Q q^{h_{R_1 R_2}(i,j)}\right)
= \prod_{k=1}^\infty \left( 1 - Q q^k\right)^k 
\prod_k \left( 1 - Q q^k\right)^{C_k(R_1, R_2)} ~.
$$
[Proof]\\
We first note that
\beqa
\sum_{i,j \geq 1} q^{h_{R_1 R_2}(i,j)} 
&=& q \left( \sum_{i=1}^\infty q^{\m_i^{R_1} - i} \right)
\left( \sum_{j=1}^\infty q^{\m_j^{R_2} -j } \right) \CR
&=& q e_1^{R_1}(q) e_1^{R_2}(q) = \widetilde f_{R_1 R_2} (q)~.
\eeqa
Therefore
\beqa
\sum_{i,j \geq 1} \log\left( 1 - Q q^{h_{R_1 R_2}(i,j)}\right) 
&=& - \sum_{i,j \geq 1} \sum_{n=1}^\infty \frac{Q^n}{n} (q^{n})^{ h_{R_1 R_2}(i,j)} = - \sum_{n=1}^\infty \frac{Q^n}{n} \widetilde f_{R_1 R_2} (q^n) \CR
&=& - \sum_{n=1}^\infty \frac{Q^n}{n} \frac{q^n}{(q^n -1)^2}
         - \sum_{n=1}^\infty \frac{Q^n}{n} \sum C_k(R_1, R_2) q^{kn} \CR
&=& - \sum_{n=1}^\infty \frac{Q^n}{n} \frac{q^n}{(q^n -1)^2}
         + \sum C_k(R_1, R_2) \log \left( 1 - Q q^{k} \right)~. \label{one}
\eeqa
On the other hand
\beqa
\sum_{k=1}^\infty k \log \left( 1 - Q q^k \right) 
&=& - \sum_{k=1}^\infty k \sum_{n=1}^\infty \frac{ (Q q^k)^n}{n}\CR
&=& - \sum_{n=1}^\infty \frac{Q^n}{n} 
\left( q^n \frac{d}{d q^n} \sum_{k=1}^\infty q^{kn} \right)\CR
&=& - \sum_{n=1}^\infty \frac{Q^n}{n} \frac{q^n}{(1-q^n)^2}~. \label{two}
\eeqa
Combining (\ref{one}) and (\ref{two}), we obtain the lemma.

Let us now proceed to Proposition 3. Our definition of $L_{R_1 R_2}(Q_1, Q_2)$ is
\beq
L_{R_1 R_2}(Q_1, Q_2) := \sum_{R_1^\prime, R_2^\prime} 
Q_1^{\ell_{R_1^\prime}} Q_2^{\ell_{R_2^\prime}}
W_{R_1 R_1^\prime} W_{R_1^\prime R_2^\prime}W_{R_2^\prime R_2}
(-1)^{\ell_{R_1^\prime}+\ell_{R_2^\prime}}
q^{-\frac{1}{2}(\k_{R_1^\prime} + \k_{R_2^\prime})}~.
\eeq
If we use the symmetric form (\ref{symmetric}) for $W_{R_1^\prime R_2^\prime}$, we find
that the factors involving $\k_{R_i}$ cancel nicely and we have
\beqa
L_{R_1 R_2}(Q_1, Q_2) &=& W_{R_1}(q) W_{R_2}(q) 
\sum_{R_1^\prime, R_2^\prime, R_3^\prime} (- Q_1)^{\ell_{R_1^\prime}}  (- Q_2)^{\ell_{R_2^\prime}} \CR
& &~~~~\times s_{R_1^\prime}(q^{\m^{R_1} + \rho}) s_{R_2^\prime}(q^{\m^{R_2} + \rho})
s_{R_1^{\prime t}/R_3^\prime} (q^\rho) s_{R_2^{\prime t} /R_3^\prime} (q^\rho) \CR
&=& W_{R_1}(q) W_{R_2}(q) 
\sum_{R_1^\prime, R_2^\prime, R_3^\prime} 
s_{R_1^\prime}(- Q_1 q^{\m^{R_1} + \rho}) s_{R_2^\prime}(- Q_2 q^{\m^{R_2} + \rho}) \CR
& &~~~~\times s_{R_1^{\prime t} /R_3^\prime} (q^\rho) s_{R_2^{\prime t} /R_3^\prime} (q^\rho)~. \label{L12}
\eeqa
Now we invoke the formulas generalizing (\ref{schur1}) and (\ref{schur2}) \cite{Mac};
\beqa
\sum_R s_{R/R_1}(x) s_{R/R_2}(y) &=& \prod_{i,j \geq 1} (1- x_i y_j)^{-1}
\sum_Q s_{R_2/Q}(x) s_{R_1/Q}(y)~, \label{schur3} \\
\sum_R s_{R/R_1^t}(x) s_{R^t/R_2}(y) &=& \prod_{i,j \geq 1} (1+ x_i y_j) 
\sum_Q s_{R_2^t/Q}(x) s_{R_1/Q^t}(y)~.  \label{schur4} 
\eeqa
Note that $R_1$ and $R_2$ are exchanged in the right hand side.
If we apply  (\ref{schur4}) for summations over $R_1^\prime$ and $R_2^\prime$
in ({\ref{L12}), only the trivial representation survives in the sum of the right hand side
of  (\ref{schur4}).
Thus we find
\beqa
L_{R_1 R_2}(Q_1, Q_2) &=& W_{R_1}(q) W_{R_2}(q) 
\prod_{i,j \geq 1} \left( 1 - Q_1 q^{\m_i^{R_1} -i -j +1} \right)
\left( 1 - Q_2 q^{\m_i^{R_2} -i -j +1} \right) \CR
& &~~~~~ \times \sum_{R^\prime_3}  s_{R^{\prime t}_3} (- Q_1 q^{\m^{R_1} + \rho})
s_{R^{\prime t}_3} (- Q_2 q^{\m^{R_2} + \rho})~.
\eeqa
Using the formula (\ref{schur2}) again, we obtain
\beqa
L_{R_1 R_2}(Q_1, Q_2) &=& W_{R_1}(q) W_{R_2}(q) 
\prod_{i,j \geq 1} \frac{\left( 1 - Q_1 q^{\m_i^{R_1} -i -j +1} \right)
\left( 1 - Q_2 q^{\m_i^{R_2} -i -j +1} \right)}
{\left( 1 - Q_1 Q_2 q^{\m_i^{R_1} -i + \m_j^{R_2} -j +1} \right)}\CR
&=& W_{R_1}(q) W_{R_2}(q) 
\prod_{k=1}^\infty \frac{(1- Q_1 q^k)^k(1- Q_2 q^k)^k}{(1 - Q_1 Q_2 q^k)^k} \CR
& &~~~ \times \prod_{k} \frac{\left( 1 - Q_1 q^k\right)^{C_k(R_1)}
\left( 1 - Q_2 q^k\right)^{C_k(R_2)}}
{\left( 1 - Q_1 Q_2 q^k \right)^{C_k(R_1, R_2)}}~.
\eeqa
This completes our proof of Proposition 3.

\bigskip

\section*{Appendix D : From GV invariants to SW prepotential}
\renewcommand{\theequation}{D.\arabic{equation}}\setcounter{equation}{0}
\renewcommand{\thesubsection}{D.\arabic{subsection}}\setcounter{subsection}{0}

Let us summarize how one can obtain the SW prepotential of
4D $SU(2)$ pure Yang-Mills theory
from the GV invariants of local Hirzebruch
surface.

At genus $g$ topological string amplitude that wraps the base ${\bf P}^1$
 $k$ times has the following singularity structure \cite{IK-P1};
\beq
f_g^{(k)}(Q) = \frac{P_g^{(k)}(Q)}{(1-Q)^{2g+4k-2}}~,
\eeq
where $Q:=Q_F=e^{-t_F}$ is the K\"ahler moduli of the fiber ${\bf P}^1$ and
the Taylor expansion of $f_g^{(k)}(Q)$ gives the GV invariants.
${P_g^{(k)}(Q)}$ is regular at $Q=1$.

The limit which reproduces four dimensional Seiberg-Witten theory
is as follows;
\beq
Q_B = \left( R\Lambda \right)^4~, \quad Q_F = e^{-4Ra}~, \quad
q = e^{-2R\hbar}
\eeq
and $R \to 0$. The terms in the topological string amplitude that survive
in this limit are
\beq
\Lambda^{4k} \sum_{g=0}^\infty \frac{1}{\hbar^{2-2g}} 
\frac{P_g^{(k)}(1)}{2^{2g-2+8k} a^{2g-2+4k}} 
= a^2 \left( \frac{\Lambda}{a} \right)^{4k}  \sum_{g=0}^\infty \frac{1}{\hbar^{2-2g}} 
\frac{P_g^{(k)}(1)}{2^{2g-2+8k} a^{2g}}~.
\eeq

Now we list the function $P_g^{(k)}(Q)$ for ${\bf F}_m$ $(m=0,1,2)$ up to instanton number three;

\subsection{$g=0$}

\begin{enumerate}

\item{${\bf F}_0$}
\beqa
P_0^{(1)} &=& 2~, \CR
P_0^{(2)} &=& \frac{2Q(3Q^2 + 4Q +3)}{(Q+1)^2}~, \\
P_0^{(3)} &=& \frac{2Q(4Q^6 + 23Q^5 + 50Q^4 + 62Q^3 + 50Q^2 + 23Q +4)}
{(Q^2 + Q +1)^2}~. \nonumber
\eeqa

\item{${\bf F}_1$}
\beqa
P_0^{(1)} &=& -(Q+1)~, \CR
P_0^{(2)} &=& \frac{2Q^2(3Q^2 + 4Q +3)}{(Q+1)^2}~, \\
P_0^{(3)} &=& \frac{-Q^3(27Q^5 + 70Q^4 + 119Q^3 + 119Q^2 + 70Q +27)}
{(Q^2 + Q +1)^2}~. \nonumber
\eeqa

\item{${\bf F}_2$}
\beqa
P_0^{(1)} &=& Q^2 + 1~, \CR
P_0^{(2)} &=& \frac{2Q^3(3Q^2 + 4Q +3)}{(Q+1)^2}~, \\
P_0^{(3)} &=& \frac{2Q^4(4Q^6 + 23Q^5 + 50Q^4 + 62Q^3 + 50Q^2 + 23Q +4)}
{(Q^2 + Q +1)^2}~. \nonumber
\eeqa
\end{enumerate}

Up to sign flip at odd instanton numbers of ${\bf F}_1$, we obtain an universal
results;
\beq
P_0^{(1)}(Q=1) = 2~, \quad P_0^{(2)}(Q=1) = 5~, \quad P_0^{(3)} (Q=1) = 48~,
\eeq
which give the coefficients of genus zero SW prepotential
\beq
{\cal F}_1 = \frac{1}{2^5}~, \quad 
{\cal F}_2 = \frac{5}{2^{14}}~, \quad 
{\cal F}_3 = \frac{3}{2^{18}}~.
\eeq

\subsection{$g=1$}

\begin{enumerate}

\item{${\bf F}_0$}
\beqa
P_1^{(1)} &=& 0~, \CR
P_1^{(2)} &=& \frac{Q^2(9Q^2 + 14Q +9)}{(Q+1)^2}~, \\
P_1^{(3)} &=& \frac{4Q^2(17Q^6 + 84Q^5 + 173Q^4 + 220Q^3 + 173Q^2 + 84Q +17)}
{(Q^2 + Q +1)^2}~. \nonumber
\eeqa

\item{${\bf F}_1$}
\beqa
P_1^{(1)} &=& 0~, \CR
P_1^{(2)} &=& \frac{Q^3(9Q^2 + 14Q +9)}{(Q+1)^2}~, \\
P_1^{(3)} &=& \frac{-2Q^3(5Q^7 + 94Q^6 + 254Q^5 + 415Q^4 + 415Q^3
+ 254Q^2 + 94Q +5)}
{(Q^2 + Q +1)^2}~. \nonumber
\eeqa

\item{${\bf F}_2$}
\beqa
P_1^{(1)} &=& 0~, \CR
P_1^{(2)} &=& \frac{Q^4(9Q^2 + 14Q +9)}{(Q+1)^2}~, \\
P_1^{(3)} &=& \frac{4Q^5(17Q^6 + 84Q^5 + 173Q^4 + 220Q^3 + 173Q^2 + 84Q +17) }
{(Q^2 + Q +1)^2}~. \nonumber
\eeqa
\end{enumerate}

Again we observe a sign flip at odd instanton numbers of ${\bf F}_1$.
we find an universal value;
\beq
P_1^{(1)}(Q=1) = 0~, \quad P_1^{(2)}(Q=1) = 8~, \quad P_1^{(3)} (Q=1) = 3072/9 = 2^{10}/3~,
\eeq
and the coefficients of genus one SW prepotential
\beq
{\cal F}_1 = 0~, \quad 
{\cal F}_2 = \frac{1}{2^{13}}~, \quad 
{\cal F}_3 = \frac{1}{3 \cdot 2^{14}}~.
\eeq

\subsection{$g=2,3$}

We have computed $P_g^{(k)}$ for ${\bf F}_0$ and 
found a precise agreement with the results reported in \cite{KMT, Hos}

\begin{enumerate}

\item{$g=2$}
\beqa
P_2^{(1)} &=& 0~, \CR
P_2^{(2)} &=& \frac{4Q^3(3Q^2 + 5Q +3)}{(Q+1)^2}~, \\
P_2^{(3)} &=& \frac{2Q^2(6Q^8 + 218Q^7 +
937Q^6 + 1868Q^5 + 2366Q^4 + 1868Q^3 + 937Q^2 + 218Q +6)}
{(Q^2 + Q +1)^2}~. \nonumber
\eeqa

\item{$g=3$}
\beqa
P_3^{(1)} &=& 0~, \CR
P_3^{(2)} &=& \frac{Q^4(15Q^2 + 26Q +15)}{(Q+1)^2}~, \\
P_3^{(3)} &=& \frac{4Q^3(39Q^8 + 624Q^7 + 2379Q^6 
+ 4616Q^5 + 5780Q^4 + 4616Q^3 + 2379Q^2 + 624Q +39)}
{(Q^2 + Q +1)^2}~. \nonumber
\eeqa

\end{enumerate}

We find
\beq
P_2^{(1)}(Q=1) = 0~, \quad P_2^{(2)}(Q=1) = 11~, \quad 
P_2^{(3)} (Q=1) = 16848/9 = 2^4\cdot3^2\cdot13~,
\eeq
and 
\beq
{\cal F}_1 = 0~, \quad 
{\cal F}_2 = \frac{11}{2^{18}}~, \quad 
{\cal F}_3 = \frac{3^2\cdot13}{2^{22}}~.
\eeq
for genus two prepotential  and for genus three
\beq
P_3^{(1)}(Q=1) = 0~, \quad P_3^{(2)}(Q=1) = 14~, 
\quad P_3^{(3)} (Q=1) = 84384/9 = 2^5\cdot293~,
\eeq
and 
\beq
{\cal F}_1 = 0~, \quad 
{\cal F}_2 = \frac{7}{2^{19}}~, \quad 
{\cal F}_3 = \frac{293}{2^{23}}~.
\eeq


\end{document}